\documentclass[aps,pra,10pt,superscriptaddress,twocolumn,floatfix,nofootinbib]{revtex4-2}
\usepackage{graphicx}
\usepackage{orcidlink}
\usepackage{amsmath,amssymb,bm,dsfont}
\usepackage{color}
\usepackage{mathrsfs}
\usepackage{setspace}
\usepackage{hyperref}
\usepackage{tikz}
\usetikzlibrary{arrows.meta,calc,decorations.pathmorphing}
\newcommand{\ket}[1]{\left| #1 \right\rangle}
\newcommand{\bra}[1]{\left\langle #1 \right|}

\usepackage[capitalise]{cleveref}
\crefname{equation}{Eq.}{Eqs.}
\crefname{figure}{Fig.}{Figs.}
\crefname{section}{Sec.}{Secs.}

\def\beq{\begin{equation}}
\def\eeq{\end{equation}}
\def\beqn{\begin{align}}
\def\eeqn{\end{align}}

\tikzset{
  photonA/.style={magenta!85!black, line width=1.1pt},
  photonB/.style={green!55!black, line width=1.1pt, dash pattern=on 1.2pt off 1.6pt},
  causalbox/.style={draw, line width=1pt, minimum size=20pt, inner sep=4pt},
  causalarrow/.style={-{Latex[length=3mm,width=2.2mm]}, line width=1.3pt},
  pics/detector/.style 2 args={code={
    \draw[fill=white, line width=0.9pt] (-0.38,0) arc (180:0:0.38) -- cycle;
    \node at (0,0.17) {\footnotesize #1};
  }},
  pics/fixedmirror/.style={code={\fill[black] (-0.34,-0.055) rectangle (0.34,0.055);}},
  pics/switchmirror/.style={code={
    \fill[black!55] (-0.3,-0.055) rectangle (0.3,0.055);
    \draw[{Latex[length=1.5mm]}-{Latex[length=1.5mm]}, black!55, line width=0.7pt] (-0.55,0) -- (-0.34,0);
  }},
  pics/bsplitter/.style={code={\draw[densely dotted, line width=0.9pt, draw=gray!95, fill=gray!25] (-0.34,-0.075) rectangle (0.34,0.075);}},
}

\begin{document}
\title{The Quantum Eraser Paradox}
\author{Colm Bracken}
\affiliation{Dept of Experimental Physics, Maynooth University, Maynooth, Co. Kildare, Ireland}
\affiliation{Astronomy \& Astrophysics Section, School of Cosmic Physics, Dublin Institute for Advanced Studies, Fitzwilliam Place, Dublin 2, D02 XF86, Ireland}
\author{Jonte R. Hance\,\orcidlink{0000-0001-8587-7618}}
\email{jonte.hance@newcastle.ac.uk}
\affiliation{Quantum Group, School of Computing, Newcastle University, 1 Science Square, Newcastle upon Tyne, NE4 5TG, UK}
\author{Sabine Hossenfelder}
\affiliation{Quantum Group, School of Computing, Newcastle University, 1 Science Square, Newcastle upon Tyne, NE4 5TG, UK}

\date{\today}

\begin{abstract}
The Delayed-Choice Quantum Eraser experiment is commonly interpreted as implying that in quantum mechanics a choice made at one time can influence an earlier event. We here suggest an extension of the experiment that results in a paradox when analysed in a local realist interpretation combined with backward causation (``dynamical retrocausality''). We argue that resolving the paradox requires giving up the idea that, in quantum mechanics, a choice can influence the past in this way, and that it instead requires a violation of Statistical Independence without (what most people think of as) retrocausality. Finally, we propose an implementation of the experiment that we believe to be possible with existing technology. This new experiment can distinguish between different types of hidden-variables theories in a way that Bell-type tests cannot: unlike in a Bell test, the measurement setting here is set by an earlier outcome, which makes the consistency of the backwards influence itself testable. We classify the fixed-point (consistency-enforcing) hidden-variables models of the experiment, which can reproduce quantum mechanics only if the backwards influence has no observable effect.
\end{abstract}

\maketitle

\section{Introduction}

In the famous Quantum Eraser experiment \cite{Scully1982Eraser}, an interference pattern can be re-created by erasing which-way information. Even more remarkably, this information can be successfully erased {\emph{after}} the which-way information was already imprinted on the state \cite{Kim2000Delayed,Walborn_2002,Scarcelli_2007,Ma1221,rivlin2022quantum,Dieguez2022Realism,KimHam2023Coherent}. This Delayed-Choice Quantum Eraser is frequently interpreted as showing the possibility of an influence on the past \cite{wheeler1978past,Ma2016DelayedReview}.

In \cite{Hance2021Ensemble}, it was argued that from the $\psi$-ensemble perspective this retrocausal interpretation makes no sense. Instead, if one wants to explain the outcome in a local realist framework, then the information about what measurement would take place must have been available already at the time of preparation. In a $\psi$-ensemble, the wave-function is not itself fundamental; it merely describes a collection of ontic states in an underlying theory.

We here propose an extension of this Delayed-Choice Quantum Eraser that can rule out local realist versions of retrocausality. By local realist theory, we here mean a hidden-variables theory that fulfils Bell's criterion of local causality (see \cite{Adlam2023Taxonomy} for the meaning of these terms). Bell's theorem \cite{Bell1964OnEPR,Bell2004Speakable,Norsen2011Bell} requires that any such local realist theory must violate one of the other assumptions of the theorem to violate Bell inequalities (as has been done experimentally \mbox{\cite{Aspect_1,GiustinaLoophole2015,Hensen2015loophole}}). We will here consider the case where Statistical Independence is violated.
This is possible through two options. The first involves assuming that the information about the measurement settings at detection is contained in the state of the hidden variables at preparation. This option is known as superdeterminism\footnote{Of which supermeasured theories \cite{Hance2022Supermeasured} are a subset.}~\cite{Hossenfelder2020Rethinking,Hossenfelder2020Perplexed,tHooft2016CA}. The other option is that the measurement setting at detection influences the initial state, which is known as retrocausality. In the terminology of the taxonomy given in~\cite{Adlam2023Taxonomy}, the retrocausal models we aim to rule out are those which allow \emph{retrocausal signalling}: models in which observables at a given time are correlated with controllable inputs (here, the choice of measurement setting) from later times.

Given both of these options violate the assumptions of Bell's theorem (do not fulfil Statistical Independence, also called measurement independence), they cannot be ruled-out (or even distinguished) with Bell-type tests. Randomising the settings, even with light from distant quasars \cite{Handsteiner2017Cosmic,Rauch2018Cosmic}, constrains when correlations between settings and hidden variables could have been established, but cannot exclude them. The test proposed here can discriminate between dynamical retrocausality and superdeterminism. What separates it from a Bell test is that the feedback loop we introduce makes one measurement setting endogenous: the setting is a function of an earlier measurement outcome. In a Bell test the settings are external inputs, and a violation of Statistical Independence is untestable by construction; with an endogenous setting, the backwards influence must act consistently with the outcome that produced the setting, and that consistency can be checked.

We want to emphasise that this type of retrocausality does \emph{not} use the notion of causality which is common in many parts of physics, in which, of two causally related events, one is the cause of the other if it is in the past light cone of the other. According to this notion of causality, retrocausality does not exist by definition \cite{ben2007impossibility}. Instead, to analyse experiments in the foundations of quantum mechanics, it has become common to use the notion of causality derived from causal diagrams \cite{Wood2015FineTuning, Pearl2009Causality,Spirtes2000Causation}, usually referred to as ``interventionist causality''. The relation between these two notions of causality has yet to be entirely clarified. For a recent discussion see \cite{Schmid2020Unscrambling}.

The extension we propose builds on the experimental design suggested by one of us in \cite{bracken2022delayed}. We argue that it supports the hypothesis that Statistical Independence is violated but that this form of retrocausality is either internally inconsistent or, when internal consistency is enforced, becomes inconsistent with observation.

A variety of retrocausal approaches to quantum mechanics have been put forward in \cite{Corry2015Retrocausal,Costa1977TimeI,Costa1979TimeII,Costa1985Concerning,Costa1987Zigzagging,Dowe1992Asymmetry,Dowe1996Backwards,Dowe1997Defense,Evans2015Retrocausality,Price1984AffectingPast,Price1994Neglected,Price1996Time,Price2008ToyModel,Price2012SymmImpliesRetrocausality,Price2015Disentangling,Sen2019Retrocausal,Sutherland1983BackwardsinTime,Sutherland1998Density,Sutherland2017HowRetroHelps,Almada2015Retrocausal,Wharton2018NewRetro,Cohen2020RetrocausalI,Cohen2020RetrocausalII,Leifer2017TimeSymmetric,Wharton2020Colloquium}. However, most of these are what Adlam refers to as ``all-at-once'' retrocausal~\cite{adlam2022two}, and so are nonlocal. We instead focus on a common feature that all local realist explanations which exploit retrocausality (dynamical retrocausal models in Adlam's terms) must share. Our argument differs from previously made ones~\cite{Arntzenius1994-ARNSC,maudlin2011quantum,berkovitz2011explanation} in that we propose a concrete experimental setup that can be realised in the near future.

This paper is organised as follows. After recalling how the Quantum Eraser works in \Cref{sec:summary}, and exploring why it is puzzling in \Cref{sec:interpret}, we will in \Cref{sec:paradox} propose an extension of the experiment that, if one believes the retrocausal explanation, results in a causality paradox. In \Cref{sec:resolve}, we will discuss what the outcome of such an experiment would be. In \Cref{sec:fixedpoint} we classify the consistency-enforcing hidden-variables models of the experiment. In \Cref{sec:extensions} we describe extensions: scans of the switching delay and of the measurement strength, an iterated version of the feedback, and a qubit-circuit implementation. In \Cref{sec:feasible} we discuss the feasibility of the experiment.

The reader who is familiar with the Quantum Eraser experiment, Bell's theorem, and the Elitzur-Vaidman bomb experiment \cite{Elitzur1993Bomb}, can jump right to \Cref{sec:paradox}.

\section{The Quantum Eraser}
\label{sec:summary}

\begin{figure}
    \centering
\includegraphics[width=0.8\linewidth]{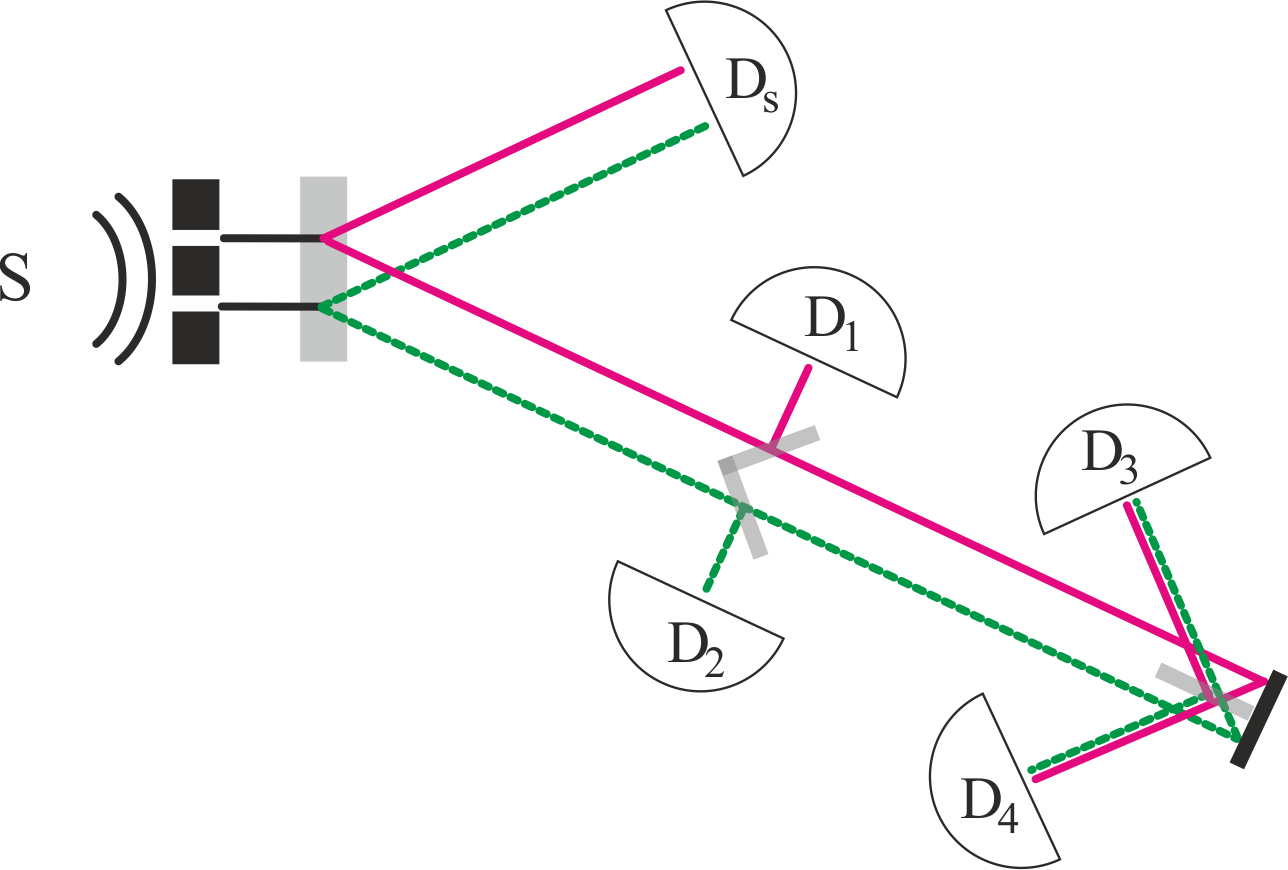}
    \caption{Experimental setup for the Quantum Eraser (based on \cite{Kim2000Delayed}). Single photons are produced at the source (S) and sent through a double slit (black), then each photon is converted into an entangled pair at a nonlinear optical crystal (grey rectangle). The photons on the upper path (solid/pink and dashed/green) go to a screen/position-resolved detector (D$_{\rm s}$). The photons on the lower path are either reflected by the switchable mirrors (dark grey bars, marked by double-headed arrows) to the first pair of detectors (D$_{\rm 1/2}$), which reveal the which-way information, or, when the switchable mirrors are removed, pass on via the fixed mirrors (solid black bars) to the beam splitter (light grey bar with dotted outline), which erases the which-way information, and are detected at the second pair of detectors (D$_{\rm 3/4}$).
    }
    \label{fig:experiment}
\end{figure}

We here briefly summarise the experiment as first realised in \cite{Kim2000Delayed}, which is close to the original proposal from \cite{Scully1982Eraser} (see Fig.~\ref{fig:experiment}). Photons are emitted from a source (S) and sent through a double slit (black). After the double-slit, they hit a nonlinear optical crystal (grey). Each photon incident on this crystal undergoes spontaneous parametric down-conversion, creating an entangled photon pair (solid/pink and dashed/green lines). Each run involves a single down-conversion event: one photon passes the double-slit and is converted into one pair, in a superposition of being on the pink (solid) and the green (dashed) paths.

One of the photons in each entangled pair is sent directly towards a detector screen (D$_{\textrm s}$), via the upper path in Fig.~\ref{fig:experiment}.
The other photon in each entangled pair travels down the lower path in Fig.~\ref{fig:experiment}, and is either reflected by the switchable mirrors to detectors D$_1$ and D$_2$, which reveal where the photon came from (the ``which-way information''), or, when the switchable mirrors are removed, passes on via the fixed mirrors to a beam splitter, which erases any which-path information, and is then measured at detectors D$_3$ and D$_4$.
Coincident detection pairs each detection on the screen with the detection at one of D$_1$--D$_4$ from the same entangled pair; optical delays ensure the lower-arm photon takes the same time to reach any of D$_1$--D$_4$, matching the upper-arm photon's time to the screen.
The question is whether or not the photons arriving at the screen D$_\textrm{s}$, coincident with a photon arriving at one of the detectors D$_\textrm{1}$--D$_\textrm{4}$, will create a two-slit interference pattern.

Theoretically one can understand the setup as follows. Let $\psi_{1}(x)$ and $\psi_{2}(x)$ denote the amplitudes at position $x$ on the screen for a photon which passed through the first and the second slit, respectively. In the absence of the crystal, the wave-function arriving at the screen would be
    \begin{equation}
        \ket{\Psi} = \frac{1}{\sqrt{2}} \int {\rm d}x \left[ \psi_1(x) + \psi_2(x) \right] \ket{x},
    \end{equation}
where $\ket{x}$ is the state of a photon at position $x$ on the screen. The probability of detecting a photon at $x$ is then $P(x) \propto |\psi_1(x)+\psi_2(x)|^2$. The two contributions are non-orthogonal, and their cross-terms give rise to an interference pattern.

However, creating an entangled pair from each incident photon at a particular location has the result of imprinting information about the position of the photon on the wave-function. We can write this as
\begin{equation}
\ket{\Psi} = \frac{1}{\sqrt{2}} \int {\rm d}x \Big[ \psi_{1}(x) \ket{{\rm D}_1} + \psi_{2}(x) \ket{{\rm D}_2} \Big] \ket{x}, \label{eq:psi}
\end{equation}
where the kets $\ket{{\rm D}_1}$ and $\ket{{\rm D}_2}$ represent the entangled partner photon going to detector D$_1$ and D$_2$, respectively. States $\ket{{\rm D}_1}$ and $\ket{{\rm D}_2}$ are orthogonal because the locations are spatially separate.

Given these two components of the wave-function are orthogonal, they cannot interfere at the screen: the orthogonal which-way information $|{\rm D}_1\rangle$ and $|{\rm D}_2\rangle$ imprinted on them prohibits interference between the components. This means the distribution of photons on the screen will be given by $P(x) \propto |\psi_{1}(x)|^2 + |\psi_{2}(x)|^2$: it will be a combination of the diffraction patterns of each slit separately, with slightly off-set peaks, and no double-slit interference.

The eraser works by instead using a measurement that projects the state (\ref{eq:psi}) on antisymmetric and symmetric combinations of the which-way information $|{\rm D}_1\rangle$ and $|{\rm D}_2\rangle$,
\begin{eqnarray}
|{\rm D}_3\rangle &=& \frac{1}{\sqrt{2}} \left( |{\rm D}_1 \rangle -  |{\rm D}_2 \rangle \right) \nonumber \\
|{\rm D}_4\rangle &=& \frac{1}{\sqrt{2}} \left(  |{\rm D}_1 \rangle + |{\rm D}_2 \rangle \right) ~. \label{34}
\end{eqnarray}
The states $|{\rm D}_3\rangle$ and $|{\rm D}_4\rangle$ also constitute an orthogonal measurement basis, and it's this basis which is measured at D$_3$ and D$_4$, respectively.

In this basis, the state (\ref{eq:psi}) takes the form
\begin{eqnarray}
\frac{1}{2} \int {\rm d}x \Big[ \left( \psi_{1} - \psi_{2}\right) |{\rm D}_3 \rangle +  \left( \psi_{1} + \psi_{2} \right) |{\rm D}_4 \rangle \Big] \ket{x} ~, \label{eq:int}
\end{eqnarray}
where the $x$-dependence of $\psi_1$ and $\psi_2$ is left implicit. The unconditioned distribution of photons on the screen is unchanged,
\begin{equation}
P(x) \propto \frac{1}{2}\left( |\psi_1+\psi_2|^2 + |\psi_1-\psi_2|^2 \right) = |\psi_1|^2 + |\psi_2|^2, \nonumber
\end{equation}
as it must be, given that we have merely rewritten the state in a different basis for the lower-path photon. However, the distributions conditioned on the outcome of the lower-path measurement, $P(x|{\rm D}_3) \propto |\psi_1-\psi_2|^2$ and $P(x|{\rm D}_4) \propto |\psi_1+\psi_2|^2$, each show interference: if we project on either $|{\rm D}_3\rangle$ or $|{\rm D}_4\rangle$ by measuring the entangled partner, the terms from the two slits can interfere.

Expression~(\ref{eq:int}) also shows that the two interference patterns, created by projecting on $|{\rm D}_3\rangle$ or $|{\rm D}_4\rangle$, respectively, are not identical. Because of the minus sign in the first term, the interference patterns are phase-shifted relative to each other. If you add them together, they re-create the original non-interference pattern. This is as we would expect, given we just changed the basis in which we measured the photons in the lower path, rather than directly transforming the state of the photons on the upper path.

For the Delayed-Choice Quantum Eraser ({\sc DCQE}), one makes the lower path long enough that the photons on this path are detected after the upper-path photon hits the screen. (The extra travel time along the added length is subtracted from the time-tags of D$_1$--D$_4$, so coincident detection still identifies the pairs.) An interference pattern is still observed in the distribution of photons on the screen conditioned on a coincident detection at D$_3$, and likewise in the distribution conditioned on a coincident detection at D$_4$.

The measurement outcome on the screen does not depend on the settings on the other path: one can merely split the photons into two different groups in two different ways \cite{ellerman2015delayed,Kastner2019,Fankhauser2019Taming,Qureshi2020Demystifying,Qureshi2021Delayed,Dieguez2022Realism,Badhan2026Standard}.

If we measure the photon on the lower path using detectors D$_1$ and D$_2$, then the entangled partners of photons reaching D$_1$ do not create a two-slit interference pattern on D$_{\rm s}$, and neither do the entangled partners of photons detected at D$_2$. If we instead measure the photon on the lower path using detectors D$_3$ and D$_4$, then photons detected on the screen coincident with a detection at D$_3$ create a two-slit interference pattern, as do those detected coincident with a detection at D$_4$. But together they do not. This is also the case in versions of the experiment using polarising plates to imprint the which-way information, as in \cite{Walborn_2002}, though the symmetric and anti-symmetric combinations differ by a complex phase.

In quantum mechanics, the outcome does not depend on how long after one photon arrives at the screen its entangled partner is measured. These measurements can even be space-like separated \cite{Ma1221}. If the pattern on D$_{\rm s}$ did depend on the measurement settings (D$_{1/2}$ or D$_{3/4}$), this could be used to violate the no-signalling theorem \cite{Shimony1993NoSig}. Therefore, the unconditioned interference pattern, made up of all detections at D$_\text{s}$ (ignoring which detector clicked at the same time as each detection), has to be independent of whether one measures the which-way information or not. The eraser effect only appears when we compare the interference patterns of conditioned subsets of these detections at D$_\text{s}$.

The challenge for this experiment is to explain how it can be that one recovers the interference pattern after it has seemingly been destroyed by imprinting the which-way information on the wave-function.

\section{Interpretation}
\label{sec:interpret}

First, the measurement results of the {\sc DCQE} experiments can be understood perfectly well in the Copenhagen interpretation. If one is willing to accept a non-local update of the wave-function upon measurement, then there is nothing puzzling about the experiment. The wave-function simply propagates until it hits the detectors, and then it `collapses' to a localised state with a probability given by Born's rule.

However, many people find the {\sc DCQE} results more disturbing (weird, strange, counter-intuitive) than, say, those of Bell-type experiments. We believe this is because a Bell-type experiment can be visualised (mentally or actually) by a local process. One simply creates two entangled particles that one can think of as little balls, which propagate in two different directions and are then detected at two separate locations. It is easy to imagine that when the particles hit the detectors, that measurement merely reveals which spin or polarisation they have, and those outcomes are correlated because the particles had a common origin.

The real puzzle of a Bell-type test lies in the fact that the outcomes at the two detectors are, for certain relative orientations of the measurement settings, more strongly correlated than one would expect from a local, classical theory in which the outcome of the measurement was determined by hidden, internal properties of the particles. But this puzzle is buried in the mathematics underlying Bell's theorem and so is harder to visualise.

In experiments like the Elitzur-Vaidman bomb experiment \cite{Elitzur1993Bomb} or the {\sc DCQE}, on the other hand, it becomes visually apparent that local realist explanations are difficult to come by, because these experiments make use of superpositions in position space (particles that ``take two paths at once'').

\begin{figure}
    \centering
\includegraphics[width=0.8\linewidth]{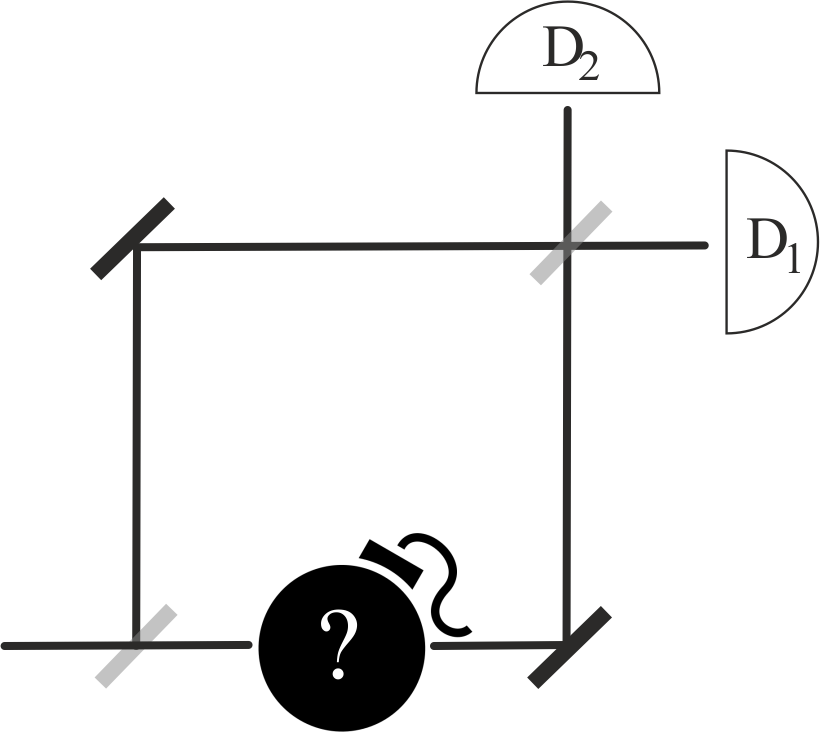}
    \caption{Sketch of the bomb experiment. Light grey bars with dotted outline are beam splitters; solid black bars are mirrors. Detector B$_2$ only registers a photon if the bomb was live and didn't explode. The experiment can thus reveal information about a counterfactual event which didn't happen.}
    \label{fig:bomb}
\end{figure}

The bomb experiment (Fig.~\ref{fig:bomb}) involves a Mach-Zehnder interferometer with a potentially faulty ``bomb'' in one path. This bomb can only be detonated by a non-demolition single-photon detection on that path. One wants to find out whether the bomb is live or a dud, ideally without blowing it up. If the photon goes along the bomb's path of the Mach-Zehnder interferometer (and the bomb works), it detonates, and the photon (and everything else) is destroyed. If the bomb is faulty, the photon travels to the merging beam splitter normally, and is always detected at B$_1$. However, if the photon goes along the other side, the bomb working changes the interference pattern. This opens the output port B$_2$, which the photon otherwise cannot reach, allowing us to test whether the bomb would have worked, without detonating it, provided the photon took the bomb-free path.
The bomb plays the role of an additional detector which reveals which-way information. If the detector is on (the bomb works) and yet doesn't click (doesn't explode), one knows the particle is on the other path.

To explain what happens in the bomb experiment, one first has to account for the destructive interference at detector B$_2$ when the bomb is a dud. In this case, by assumption, no interaction between the photon and the bomb takes place at all, and we just have a normal Mach-Zehnder interferometer. The destructive interference at B$_2$ then requires one to accept that the photon acts like a wave and takes both possible paths. But if the bomb is in place, works, and yet didn't explode, then we know the photon must have taken the path on which there is no bomb. This means, if one wants a local explanation, then the propagation of the photon must depend on the detector setting (whether the bomb works or not), before the photon has even reached the detector (the bomb).

Of course one could argue that we knew this already: a local realist hidden-variables explanation of quantum mechanical phenomena requires that the path of a particle depends on the measurement setting at the time of measurement\footnote{The measurement settings in Bell's theorem are those at the time of measurement. It does not matter at which time or how the settings were chosen, it only matters what they are when the actual measurement takes place.}. This, after all, is what Bell's theorem tells us: If we insist on a local explanation, then Statistical Independence must be violated \cite{Hossenfelder2020Rethinking}, either by superdeterminism or by retrocausality. However, the bomb experiment makes it much more apparent just what this entails. Most notably, it shows just how easy it is to explain the observations by violating Statistical Independence: If one measures the which-way information (uses a bomb that works), the photon either goes via one path or the other, but not both. Then, it has a 50:50 chance of being transmitted or reflected by the beam splitter. That's it.\footnote{In such a case, interaction-free (``counterfactual'') communication, computation, and imaging, such as that in \cite{Hosten2006CFComp,Salih2013Protocol,Salih2016Qubit,Salih2018Laws,Salih2020DetTele,Hance2021CFGI,Hance2021Quantum,Salih2021EFQubit}, is possible locally specifically because of this statistical independence violation; the message/information is to some degree encoded in the detector settings.}

The quantum eraser seems perplexing for a similar reason. If one measures the which-way information at D$_1$ and D$_2$, then a local realist explanation requires that the original photon either went through the upper slit, converting into the upper entangled pair (pink/solid line) or the lower slit, converting into the lower entangled pair (green/dashed line), but not both. If, however, we ignore the which-way information and use instead detectors D$_3$ or D$_4$, then it seems a particle going to, say, D$_3$, must have gone through both slits, so it can interfere with itself (and similarly for D$_4$). That is, either the choice of whether to use the eraser decides what happened at the screen earlier (retrocausality), or the photons' paths depended on the measurement setting all along (superdeterminism).

This argument doesn't work as well for the quantum eraser as for the bomb experiment, because, in interfering the two paths (as is done to erase the which-path information), the paths necessarily have to come together in one place. One can therefore explain the quantum eraser just by postulating that the photons always go via only one path, it's just that at the beam splitter, half of them pass through and the other half of them do not, and the positions their entangled partners go to on the screen are correlated with whether the photons at the eraser go through the beam splitter or not (see \cite{LaCour2021Classical} for an explicit local model of the eraser, built instead from classical fields and threshold detectors).

From a purely mathematical perspective, these two experiments do not teach us anything that we didn't already know from violations of Bell's inequality, namely that a local realist hidden-variables theory must be either superdeterministic or retrocausal. However, in experiments like the Elitzur-Vaidman bomb experiment and the {\sc DCQE} the difficulty becomes much more tangible because one can no longer visualise the particles as going on paths without running into difficulties.

We will in the following argue that the {\sc DCQE} experiment can be extended so that, like the bomb experiment, a local realist explanation is possible only if the path of the photon either (a) allows a retrocausal influence or (b) depends on the measurement setting all along. We will then add another twist to rule out option (a).

\section{The Paradox}
\label{sec:paradox}

\begin{figure}
    \centering
\includegraphics[width=0.8\linewidth]{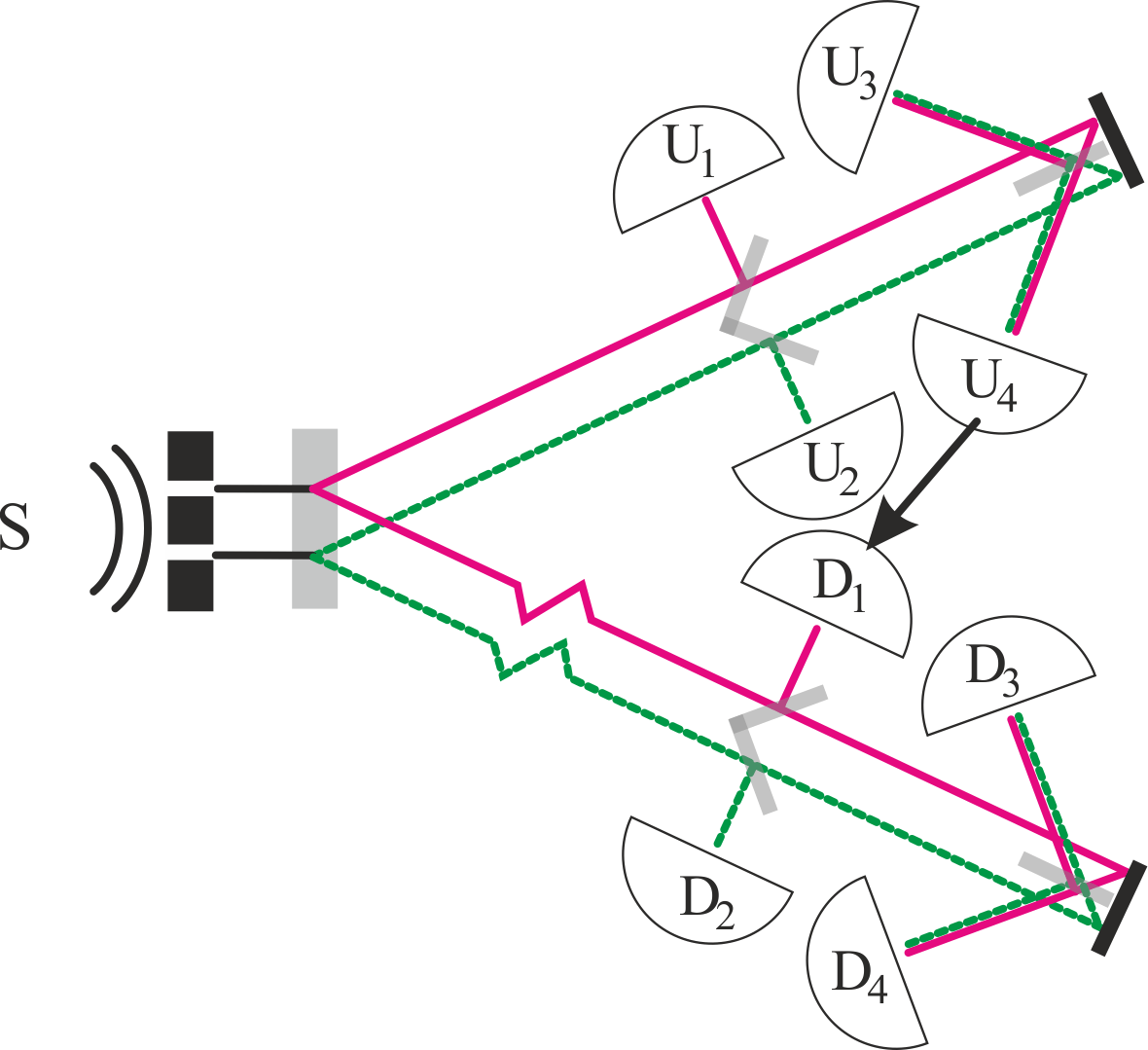}
    \caption{Delayed-choice quantum eraser with feedback (black arrow): a detection at U$_4$ switches in the mirror which sends the lower-path photon to D$_1$. Single photons are produced at the source (S) and sent through a double slit (black), then each photon is converted into an entangled pair with a nonlinear optical crystal (grey rectangle). Zig-zags in the lower two paths indicate that these paths can be much longer than the upper two. Solid black bars are fixed mirrors; dark grey bars (marked by double-headed arrows) are switchable mirrors, which can be inserted or removed; light grey bars with dotted outline are beam splitters.}
    \label{fig:screen}
\end{figure}

What goes on in the {\sc DCQE} experiment becomes clearer when we replace the interference screen D$_{\rm s}$ with another eraser quartet of detectors (see Fig.~\ref{fig:screen}, ignoring the black arrow for now). Like in the {\sc DCQE} experiment, photons are emitted from a single photon source (S) and sent through a double slit (black). After the double-slit, they hit a nonlinear optical crystal (grey) which, by spontaneous parametric down-conversion, creates an entangled pair of photons from each photon incident on the crystal (solid/pink and dashed/green lines). That is, at the down-conversion crystal, we create the entangled state
\begin{equation}
|\Psi \rangle = \frac{1}{\sqrt{2}} \Big( |{\rm U}_1 \rangle |{\rm D}_1 \rangle + |{\rm U}_2 \rangle |{\rm D}_2\rangle \Big), ~\label{eq:psi2}
\end{equation}
where each component of the wave-function is labelled by the detector it evolves towards.
On both the upper and the lower paths, we can choose whether to measure the which-way information (D$_{1/2}$, U$_{1/2}$) or whether to erase it (D$_{3/4}$, U$_{3/4}$), or some combination thereof, by inserting or removing the switchable mirrors (dark grey bars). The detectors U$_3$ and U$_4$ measure the same superpositions as defined in (\ref{34}), just with U's instead of D's. Note that, in this local realist setting, we assume measurement at a detector collapses the wave-function.

We assume that the lower paths leading to the D-detectors are much longer, so that those measurements happen later than the ones on the upper paths.

The photon pairs which emerge from the optical crystal are phase-matched; depending on the source they may also be polarisation-entangled, but we do not use the polarisation. The entanglement relevant here is in the path degree of freedom: the pairs are in one of two states, which we denote $|{\rm A}\rangle$ and $|{\rm B}\rangle$ depending on which slit they came from.

Again, we adopt the convention that the first entry in the product state denotes the photon going up, the second the photon going down. Then the state after the crystal can be written as
\begin{eqnarray}
\ket{\Psi} &=& \frac{1}{\sqrt{2}} \Big( \ket{{\rm A}}_{\rm U} \ket{{\rm A}}_{\rm D} + \ket{{\rm B}}_{\rm U} \ket{{\rm B}}_{\rm D} \Big) ~. \label{eq:psin}
\end{eqnarray}
The actions of the detectors are
\begin{eqnarray}
 {\rm U}_1:  \ket{{\rm A}}_{\rm U}  \bra{{\rm A}}_{\rm U}  \otimes {\rm \mathds{1}}_{\rm D}  ~,~
 {\rm U}_2:  \ket{{\rm B}}_{\rm U}   \bra{{\rm B}}_{\rm U}  \otimes {\rm \mathds{1}}_{\rm D}   ~, \nonumber \\
 {\rm D}_1: {\rm \mathds{1}}_{\rm U}  \otimes |{\rm A}  \rangle_{\rm D} \langle  {\rm A}|_{\rm D}  ~,~
 {\rm D}_2:  {\rm \mathds{1}}_{\rm U}  \otimes |{\rm B} \rangle_{\rm D}   \langle {\rm B} |_{\rm D}    ~.
\end{eqnarray}
The detectors U$_{3/4}$ and D$_{3/4}$ instead project on the bases
\begin{eqnarray}
 \ket{\pm}_{\rm U/D}  := \frac{1}{\sqrt{2}} \left(\ket{{\rm A}}_{\rm U/D}\pm \ket{{\rm B}}_{\rm U/D}\right)~,
\end{eqnarray}
so that
\begin{eqnarray}
{\rm U}_3:    | - \rangle_{\rm U}  \langle -|_{\rm U}  \otimes {\rm \mathds{1}}_{\rm D} ~,~
 {\rm U}_4:   |{\rm +} \rangle_{\rm U}   \langle {\rm +} |_{\rm U} \otimes {\rm \mathds{1}}_{\rm D}    ~, \nonumber \\
{\rm D}_3: {\rm \mathds{1}}_{\rm U}  \otimes |{\rm -} \rangle_{\rm D}  \langle  {\rm -}|_{\rm D}  ~,~
 {\rm D}_4:  {\rm \mathds{1}}_{\rm U}  \otimes |{\rm +} \rangle_{\rm D}   \langle {\rm +} |_{\rm D}    ~.
\end{eqnarray}
Note that since we only consider local realist settings, we assume that detectors only exist in pointer states, not in superpositions of these: a detector either detects the photon or it does not, but not both.

This modification makes the similarity to Bell-type experiments obvious. We could measure the which-way information on both sides (D$_{1/2}$, U$_{1/2}$), and find that the results are perfectly correlated. We could instead erase the which-way information on both sides (D$_{3/4}$, U$_{3/4}$) and find that the results are also perfectly correlated. Or we could measure the which-way information on one side and not on the other ((D$_{1/2}$ and U$_{3/4}$) or (D$_{3/4}$ and U$_{1/2}$)) and find the results are entirely uncorrelated. So far, so unsurprising.

If one wants to come up with a local realist explanation, then one does not have many options. We know that, since we can measure the photons at the U$_{1/2}$ and D$_{1/2}$ detectors, if they travel there locally they must have come from either slit A or B, but not both. (In quantum mechanics, of course, the wave-function is non-locally updated upon reaching the detectors, so one cannot draw this conclusion and the argument stops here.) In the local realist framework, in contrast, we are forced to conclude that if we do not use the which-path detectors, then two photons coming from A must synchronise what they do at their respective beam splitters, and so must those coming from B: one just imagines the photons are particles going via one specific path, which go through the beam splitter half of the time, with entangled photons doing the same on each of their beam splitters. This seems odd, but so far, no retrocausality or superdeterminism is required.

One thing one could do now is to create a mixture of which-way and erasure settings. That would constitute a Bell-type experiment. Then we could use violations of Bell's inequality to draw conclusions about violations of Statistical Independence. However, this setup is technically cumbersome and also teaches us nothing new.

Instead, we can make a third type of measurement, which is similar to that in the bomb experiment: Measure on the upper path with U$_3$ and U$_4$, but measure on the lower path with D$_1$, and then D$_3$ and D$_4$ (for those photons that did not go to D$_1$).
Now remember that to explain the previous measurements with a local realist visualisation, we had to settle on the case where we have a particle going along one particular path (pink or green, but not both), and at a beam splitter, it goes through only half of the time, and pairs of entangled particles do the same at their respective splitters. This explanation would tell us that in this third type of measurement, for those photons that do not appear in detector D$_1$, outcomes for their entangled partners at U$_{3/4}$ and D$_{3/4}$ are still correlated. Alas, in quantum mechanics they should now be uncorrelated.

This third measurement causes problems for local realist hidden variables for the same reason as the Elitzur-Vaidman \mbox{\cite{Elitzur1993Bomb}} bomb experiment does: It seems that the photon needs to know what type of measurement will take place before getting on its way. We note in passing that the toy model put forward in \mbox{\cite{Catani2021Interference}} does not avoid this problem. Rather, the problem does not appear in the cases considered in this reference because Catani et al assume that the which-path measurements cannot be completely destructive: e.g., for the bomb-tester, randomising the ``phase-value'' of the photon for that path rather than setting it to zero \mbox{\cite{hance2022comment,catani2022reply,hance2026noncontextualversuscontextualinterferometry}}).

To fix this problem (in a locally realist way), we can then either (a) require that turning on detector D$_1$ determines what happened at U$_{3/4}$, that is, in the past. This is the retrocausal option whose causal relations are depicted in Fig.~\ref{fig:causal01}. Alternatively, we (b) accept that the paths of the photons were dependent on the detector settings at the time of measurement.\footnote{Again, those are the detector settings {\emph{at the time of measurement}}. If the path could depend on the detector setting at an earlier time, that would be non-local, not superdeterministic.} This is the superdeterministic option whose causal relations are depicted in Fig.~\ref{fig:causal02}. This is the same conclusion one draws from Bell-type tests: a local realist hidden-variables explanation is possible only if one accepts either retrocausality or superdeterminism. Causal models of this kind for delayed-choice experiments (without the feedback we introduce below) were analysed in \cite{Chaves2018CausalDC}, where it was shown that Wheeler's original delayed-choice experiment can be reproduced by a simple classical causal model, and a modified experiment was proposed whose statistics rule out any non-retrocausal hidden-variable model of the same dimension as the quantum system it mimics; experimental tests along these lines have since been performed \cite{Polino2019DIDelayed,Yu2019CausalDelayed}.

\begin{figure}
    \centering
    \begin{tikzpicture}
    \node[causalbox] (U) at (0,2.1) {U$_{3/4}$};
    \node[causalbox] (D1) at (1.9,2.1) {D$_1$};
    \node[causalbox] (D34) at (3.8,2.1) {D$_{3/4}$};
    \node[causalbox] (S) at (1.9,0) {S};
    \draw[causalarrow] (D1) -- (S);
    \draw[causalarrow] (S) -- (U);
    \draw[causalarrow] (S) -- (D34);
    \end{tikzpicture}
       \caption{Retrocausality, causal relations.}
    \label{fig:causal01}
\end{figure}

\begin{figure}
    \centering
    \begin{tikzpicture}
    \node[causalbox] (U) at (0,2.1) {U$_{3/4}$};
    \node[causalbox] (D1) at (1.9,2.1) {D$_1$};
    \node[causalbox] (D34) at (3.8,2.1) {D$_{3/4}$};
    \node[causalbox] (S) at (1.9,0) {S};
    \draw[causalarrow] (S) -- (D1);
    \draw[causalarrow] (S) -- (U);
    \draw[causalarrow] (S) -- (D34);
    \end{tikzpicture}
    \caption{Superdeterminism, causal relations.}
    \label{fig:causal02}
\end{figure}

To rule out (a), we propose to use a detection at U$_4$ to turn on D$_1$, which creates the causal relations depicted in Fig.~\ref{fig:causal03}. If using detector D$_1$ indeed retrocausally determined that the entangled partner of a photon which did not go to D$_1$ must have had a 50\% chance of going to U$_3$, then in half of the runs in which the photon went to U$_4$ we create a causal paradox: the photon went to U$_4$, so we turned on D$_1$, but as a result it went to U$_3$!

\begin{figure}
    \centering
    \begin{tikzpicture}
    \node[causalbox] (U) at (0,2.1) {U$_{3/4}$};
    \node[causalbox] (D1) at (1.9,2.1) {D$_1$};
    \node[causalbox] (D34) at (3.8,2.1) {D$_{3/4}$};
    \node[causalbox] (S) at (1.9,0) {S};
    \draw[causalarrow] (U) -- (D1);
    \draw[causalarrow] (D1) -- (S);
    \draw[causalarrow] (S) -- (U);
    \draw[causalarrow] (S) -- (D34);
    \end{tikzpicture}
    \caption{Retrocausality with feedback loop, causal relations.}
    \label{fig:causal03}
\end{figure}

This feedback loop seems to realise an instance of the well-known grandfather paradox, in which you go back in time and kill your own grandfather, so you're never born and can't go back in time to kill your own grandfather. Did you, or did you not, kill your grandfather? We can ask here the same way: ``Did, or didn't, the photon go to U$_4$?'' Consistency conditions for causal loops of this kind have been studied extensively in the context of closed timelike curves \cite{Deutsch1991CTC,Lloyd2011CTC}; we return to them in \Cref{sec:fixedpoint}.

\section{Resolving the Paradox}
\label{sec:resolve}

It seems that retrocausality cannot predict what will happen in this experiment. Clearly the contradiction itself will not manifest, for what would that mean? Would the world end in a poof? Would an error message pop up, crashing the simulation that is our reality? Certainly not (we hope). But then what would happen?

In a dynamical retrocausal model, the backwards influence of turning on D$_1$ propagates from the measurement setting to the preparation, and the forward evolution is then re-run with the photons on definite paths. To extract a prediction, the re-run dynamics has to be specified. Take first the natural reading of the retrocausal story in \Cref{sec:paradox}: every U$_4$ detection triggers the re-preparation, and the re-run outcome at the upper beam splitter is a fresh $50{:}50$ draw. Then the U$_4$ branch always re-starts the loop, and the only outcome which does not is that the photon on the upper path just never goes to U$_4$. If it doesn't go to U$_4$, D$_1$ never turns on, and the result is always U$_3$ and D$_3$ together. This is internally consistent (no causal paradox occurs) but it's not what quantum mechanics predicts. Quantum mechanics predicts that, per run, the outcome pair (U$_3$, D$_3$) occurs with probability $1/2$, (U$_4$, D$_1$) with probability $1/4$, and (U$_4$, D$_3$) and (U$_4$, D$_4$) with probability $1/8$ each. A sequence of $N$ runs in which every photon goes to U$_3$ and D$_3$ therefore has probability $2^{-N}$: the longer the sequence, the less likely. This version of the dynamical retrocausal explanation can hence be ruled out experimentally.

Other specifications of the re-run dynamics give different, but still non-quantum, statistics. If the backwards influence acts only when a detection changes the setting, so that a re-run which again ends in U$_4$ terminates the loop, then iterating until the history stops changing gives $P({\rm U}_3, {\rm D}_3) = 2/3$, $P({\rm U}_4, {\rm D}_1) = 1/6$, and $P({\rm U}_4, {\rm D}_3) = P({\rm U}_4, {\rm D}_4) = 1/12$. The remaining possibility, in which the outcomes of the re-run are fixed by the hidden variables rather than drawn afresh, is the fixed-point case, which we treat in \Cref{sec:fixedpoint}.

This argument targets dynamical retrocausality. In all-at-once retrocausal models \cite{adlam2022two,Wharton2018NewRetro,Ridley2025TimeEvent}, consistency is instead imposed globally, on the whole history at once, rather than by a dynamically mediated backwards influence. In such models, runs in which the photon goes to U$_4$ can be self-consistent: the hidden variables are simply such that, with D$_1$ on, the outcome on the upper path is U$_4$. Models of this type can reproduce the quantum mechanical predictions for our experiment, but they achieve this by jointly constraining the hidden variables and the measurement settings across the whole experiment, which is a violation of the Statistical Independence assumption of Bell's theorem rather than a dynamical influence of the future on the past.\footnote{A comment \cite{Drezet2021NonParadox} on an earlier version of this argument raised a related objection: that in a deterministic retrocausal model, no paradox arises. We agree, and this is the point: once consistency is enforced, either the model deviates from the predictions of quantum mechanics (dynamical retrocausality), or the apparent backwards causation reduces to a constraint jointly on the hidden variables and settings, i.e., a violation of Statistical Independence. See also our reply \cite{Bracken2021Reply}.} \Cref{sec:fixedpoint} makes this quantitative.

If the experiment bears out the predictions of quantum mechanics, we can conclude that, in any local realist hidden-variables theory, Statistical Independence must have been violated all along. As previously discussed in \cite{Hance2021Ensemble}, this also solves the measurement problem and removes other known inconsistencies of quantum mechanics, such as the contradictions resulting from Extended Wigner's Friend scenarios \cite{Frauchiger2018,Bong2020NoGo}.

\section{Consistent Hidden-Variables Models of the Feedback Experiment}
\label{sec:fixedpoint}

We now classify the local realist hidden-variables models of the feedback experiment which resolve the loop through a global consistency condition.

Consider any local realist hidden-variables model of the feedback experiment in which each run is described by a hidden state $\lambda$, distributed with measure $\mu$, together with response functions returning the outcomes given for each setting of the switchable mirror: $u_s(\lambda) \in \{{\rm U}_3, {\rm U}_4\}$ and $d_s(\lambda)$, for $s \in \{\text{on},\text{off}\}$. (Stochastic models are included by absorbing any additional randomness into $\lambda$.) The setting is endogenous: the mirror is switched in iff the realised upper outcome is U$_4$. A run is consistent when the realised setting and the realised outcome imply each other,
\begin{equation}
s = \text{on} ~\Leftrightarrow~ u_s(\lambda) = {\rm U}_4. \label{eq:fpcond}
\end{equation}

We require that the model reproduces quantum mechanics when the same apparatus is run with externally fixed settings, with the same measure $\mu$ over $\lambda$ as in the feedback runs. (A model which drops this requirement correlates the preparation with the experimental configuration from the outset, and is superdeterministic by construction.) This means that, with the mirror out, the outcomes are perfectly correlated, $d_\text{off}(\lambda) = {\rm D}_3 \Leftrightarrow u_\text{off}(\lambda) = {\rm U}_3$, with $\mu(u_\text{off} = {\rm U}_4) = 1/2$; alternatively, with the mirror in, $\mu(u_\text{on} = {\rm U}_4) = 1/2$ and $\mu(d_\text{on} = {\rm D}_1) = 1/2$, with $u_\text{on}$ and $d_\text{on}$ independent and $d_\text{on}$ uniform on $\{{\rm D}_3,{\rm D}_4\}$ when it is not D$_1$.

Each $\lambda$ falls into one of four types, according to the pair $(u_\text{on}(\lambda), u_\text{off}(\lambda))$. Writing $p_2 := \mu(u_\text{on} = {\rm U}_4,\, u_\text{off} = {\rm U}_4)$, the static constraints fix the measures of the types up to this single parameter (see Appendix~\ref{app:fixedpoint}):
\begin{eqnarray}
\mu({\rm U}_4, {\rm U}_3) = \mu({\rm U}_3, {\rm U}_4) &=& \tfrac{1}{2} - p_2 ~, \nonumber \\
\mu({\rm U}_3, {\rm U}_3) &=& p_2 ~. \label{eq:types}
\end{eqnarray}
Checking condition \eqref{eq:fpcond} for each type, we see that $({\rm U}_4, {\rm U}_4)$ admits just one consistent history (U$_4$, mirror in), and $({\rm U}_3, {\rm U}_3)$ also admits just one (U$_3$, mirror out); $({\rm U}_4, {\rm U}_3)$ admits two consistent histories, so the model must give a rule for which is realised; and $({\rm U}_3, {\rm U}_4)$ admits none. This last type realises \Cref{sec:paradox}'s grandfather paradox at the level of an individual run.

This can lead to two different cases. If $p_2 = 1/2$, no $\lambda$ is paradoxical or ambiguous, and the model reproduces the quantum mechanical feedback statistics of \Cref{sec:resolve} (see Appendix~\ref{app:fixedpoint}). However, then $u_\text{on}(\lambda) = u_\text{off}(\lambda)$ for almost every $\lambda$, meaning the setting of the mirror has no effect on the earlier outcome in any individual run. In this case, the backwards channel is inert, there is no retrocausal signalling in the sense of \cite{Adlam2023Taxonomy}, making the scenario just that of a hidden variable correlated with the (endogenous) setting, i.e., superdeterminism. Note that this violation of Statistical Independence needs no conspiratorial initial conditions---the correlation between $\lambda$ and the setting is produced by an ordinary forward causal chain, from $\lambda$ through the U outcome to the setting. The fine-tuning critiques of superdeterministic models \cite{Sen2020SuperdetI,Sen2020SuperdetII} therefore do not apply to this model.

If instead $p_2 < 1/2$, a fraction $1/2 - p_2$ of the $\lambda$ admit no consistent history and must be excluded from the feedback ensemble, and an equal fraction admit two. As shown in Appendix~\ref{app:fixedpoint}, the model then reproduces quantum mechanics only if both the ambiguous $\lambda$ are resolved to the mirror-in history with probability $1/2$, and the lower-arm response $d_\text{on}$ is uncorrelated with the type of $\lambda$ (not merely with $u_\text{on}$, which is all the static phenomenology imposes) and with the resolution rule. Each of these is a constraint on the hidden variables which depends on the experimental configuration (a violation of Statistical Independence), fine-tuned in the sense of \cite{Wood2015FineTuning,Almada2015Retrocausal,Sen2020SuperdetI,Sen2020SuperdetII} such that the backwards influence never becomes visible in the statistics.

The feedback experiment therefore leaves consistency-enforcing models two options: reproduce quantum mechanics with an inert backwards channel (and be superdeterministic in all but name), or hide a live backwards channel behind fine-tuned reweighting of the hidden-variables measure.

The two possible outcomes of the experiment therefore teach us different lessons. If quantum mechanics is confirmed, the local realist options narrow to superdeterminism and all-at-once retrocausality, and dynamical retrocausal signalling is excluded. If the statistics deviate, we instead get positive evidence for a dynamically-mediated backwards influence. For this second case, we now propose tests which would then locate where, and how strongly, this dynamically-mediated backwards influence acts.

\section{Extensions}
\label{sec:extensions}

The experiment proposed in \Cref{sec:paradox} tests whether a dynamically-mediated backwards influence exists. We now describe three variations of this experiment (scanning the switching delay, varying the strength of the which-way measurement, and iterating the feedback) which, if there is a dynamically-mediated backwards influence, allow us to locate where and how strongly this influence acts. We also give a qubit-circuit implementation for these experiments.

\subsection{Scanning the switching delay}
\label{sec:delayscan}

In the feedback experiment, the switching signal triggered by a U$_4$ detection must reach the switchable mirror before the lower-arm photon does. Let $t_{\rm m}$ be the time at which the lower-arm wavefront passes the mirror location, $t_{\rm sw}$ the time at which the mirror has settled following a U$_4$ detection, and $\tau := t_{\rm m} - t_{\rm sw}$. Quantum mechanics predicts that the outcome statistics depend only on the state of the mirror when the wavefront passes it: for $\tau > 0$, the outcome statistics match the feedback statistics from \Cref{sec:resolve}, while for $\tau < 0$ the feedback is inoperative and the perfectly correlated statistics are recovered ((U$_3$, D$_3$) and (U$_4$, D$_4$), each with probability $1/2$), with a transition over the mirror's switching window. By inserting a variable optical delay into the lower arm, $\tau$ can be scanned through zero and this profile mapped out.

For a dynamical retrocausal model, in contrast, the backwards influence has to reach the preparation event and propagate forward again, so its effect should depend not only on the local state of the mirror at passage but on the spacetime relation between the switching event, the source, and the detectors. Scanning $\tau$ and the arm lengths independently decouples the two, and any deviation from the quantum mechanical profile would locate, in spacetime, the segment along which the backwards influence operates. This is the delayed-choice analogue of the experiment \cite{Salart2008Speed} and theoretical analyses \cite{Bancal2012Finite} which constrain hypothetical finite-speed influences in Bell tests. Fast electro-optic switching, of the type long used in delayed-choice experiments \cite{Jacques2007Wheeler,Ma1221}, makes such scanning experimentally viable, and the $\tau < 0$ region doubles as a systematic check that the feedback in the main experiment really does complete before the wavefront passes the mirror.

\subsection{Varying the measurement strength}
\label{sec:weak}

The which-way detector D$_1$ need not be projective. Let us instead replace the switchable mirror by a switchable pickoff of reflectivity $\eta$, which diverts a fraction of the amplitude in the slit-A path to D$_1$. Setting $\eta = 1$ recovers the original experiment, and $\eta = 0$ gives the mirror-out configuration. For the feedback protocol (a U$_4$ detection switches the pickoff in), quantum mechanics predicts
\begin{eqnarray}
P({\rm U}_3, {\rm D}_3) &=& \frac{1}{2}~, \qquad P({\rm U}_4, {\rm D}_1) = \frac{\eta}{4}~, \nonumber\\
P({\rm U}_4, {\rm D}_3) &=& \frac{\left(1 - \sqrt{1-\eta}\right)^2}{8}~, \nonumber\\
P({\rm U}_4, {\rm D}_4) &=& \frac{\left(1 + \sqrt{1-\eta}\right)^2}{8}~. \label{eq:etafeedback}
\end{eqnarray}
For the third measurement of \Cref{sec:paradox} with static settings, the correlator between the U$_{3/4}$ and D$_{3/4}$ outcomes among the runs in which D$_1$ did not click interpolates smoothly,
\begin{equation}
E(\eta) = \frac{2\sqrt{1-\eta}}{2-\eta}~, \label{eq:etacorr}
\end{equation}
falling from $E = 1$ at $\eta = 0$ to $E = 0$ at $\eta = 1$. The simplest synchronised-particle account of \Cref{sec:paradox}, in which photons that are not diverted to D$_1$ are undisturbed, instead keeps $E(\eta) = 1$ for all $\eta$. The scan thus replaces the correlated/uncorrelated dichotomy of the third measurement with a full curve, and lets the experimenter trade signal (the paradox-generating fraction of runs is proportional to $\eta$) against disturbance. For related variable-strength analyses of the eraser, see \cite{rivlin2022quantum,Dieguez2022Realism}.

\subsection{Iterated feedback}
\label{sec:iterated}

The feedback loop can also be iterated. Take a three-photon path-entangled state
\begin{equation}
\ket{\Psi} = \frac{1}{\sqrt{2}}\Big( \ket{{\rm A}}\ket{{\rm A}}\ket{{\rm A}} + \ket{{\rm B}}\ket{{\rm B}}\ket{{\rm B}} \Big)~, \label{eq:ghz}
\end{equation}
with arms U, D and E of increasing length, produced for example by converting one photon of the pair in a second down-conversion stage \cite{Hubel2010Triplets,Hamel2014Triplets} or, as proposed in \cite{Corona2011TOSPDC}, by direct third-order down-conversion. Let both the D and E arms have a which-way/eraser stage as in Fig.~\ref{fig:screen}. If we then measure the U photon in the $\pm$ basis, where a U$_4$ outcome switches in the which-way mirror for D$_1$, and a D$_4$ outcome (possible only when the D photon reached its eraser) switches in the which-way mirror for E$_1$, then quantum mechanics predicts
\begin{eqnarray}
P(3,3,4) &=& \tfrac{1}{4}~, \nonumber\\
P(3,4,1) = P(4,1,3) = P(4,1,4) &=& \tfrac{1}{8}~, \nonumber\\
P(3,4,e) = P(4,3,e) = P(4,4,e) &=& \tfrac{1}{16}~, \label{eq:ghzdist}
\end{eqnarray}
where $P(u,d,e) := P({\rm U}_u, {\rm D}_d, {\rm E}_e)$, the label $e \in \{3,4\}$ stands for either eraser outcome on the E arm, and all other combinations have probability zero.

A dynamical retrocausal model of the re-randomising type discussed in \Cref{sec:resolve} must now avoid re-starting either loop. The only history which switches nothing in is (U$_3$, D$_3$, E$_4$): E$_4$ is required by the three-way erased correlations, which permit only outcome triples with an even number of $-$ outcomes. Its quantum-mechanical probability is $1/4$ per run, so a sequence of $N$ runs all giving (U$_3$, D$_3$, E$_4$) has probability $4^{-N}$. Each additional feedback layer therefore doubles the discrimination rate per run: with $k$ layers, the one history which switches nothing in has probability $2^{-k}$, and $N$ runs discriminate at $(2^k)^{-N}$. \Cref{sec:fixedpoint}'s consistency accounting compounds in the same way, since the hidden state must carry consistent response functions for every combination of the $k$ settings, and the reweighting required of a model with a live backwards channel must be fine-tuned layer by layer. Photon-triplet rates are low in current sources \cite{Hubel2010Triplets,Hamel2014Triplets}, which makes a photonic implementation a longer-term goal, but the circuit implementation we give below easily extends to this sort of iterated feedback.

\section{Feasibility}
\label{sec:feasible}

By combining the bomb experiment and the quantum eraser, the experiment proposed above allows us to perform a new test of local realist theories. We will now discuss what it would take to perform this experiment.

\subsection{Using Optics}

The most practical method to create high-intensity entangled photon pairs is spontaneous parametric down-conversion ({\sc SPDC}), as demonstrated in, e.g., \cite{down_conversion_1}. Energy conservation in {\sc SPDC} requires that the sum of the energies of the two down-converted photons equal the energy of the pump photon \cite{Karab2018SPDC}. This technique typically utilises nonlinear crystals such as barium borate ({\sc BBO}) \cite{BBO_1}; earlier experiments instead used the radiative cascade of calcium atoms as a source of entangled photon pairs \cite{Aspect_1}. We will focus on the former for our purposes.

When a monochromatic light source of wavelength $\lambda_0$ illuminates the {\sc BBO} crystal, a small fraction of the pump photons are converted into pairs of down-converted photons. Because of the difficulty in producing short-wavelength lasers, and the limits on the operational wavelength range of nonlinear crystals, experiments to date \cite{Kim2000Delayed, Ma1221, Scarcelli_2007} have typically used optical or ultraviolet `pump' wavelengths of $\lambda_0 \approx 350$--$450$~nm to excite the crystals. The resulting down-converted photons are position/momentum entangled and, in the degenerate case, each has half the energy of the pump photon, and so a wavelength of $\lambda_{\rm DC} = 2\lambda_0$, which is typically in the red or infrared range (depending on the pump $\lambda_0$).

At such low photon energies, conventional detector arrays such as charge-coupled devices ({\sc CCD}s) simply cannot reliably count individual photons due to read noise: the semiconductor energy band gap is of the same order of magnitude as the energy of the photons themselves ($\approx 1$ eV). Because of this, noise contributions can be mistaken for photon events, and only a large number of photon events can generate a reliable signal. This is not ideal for photon-counting experiments such as we would like to do with the quantum eraser.

We believe the proposed measurements to be possible due to recent advances in detector technology for optical/near-infrared photons. Conveniently, the setup proposed here does not involve measuring across an interference plane or line, as is typically done in a {\sc DCQE} experiment \mbox{\cite{Kim2000Delayed, Ma1221}}. We can simply measure correlations in particular detector combinations using coincidence detection. For this, a single pixel sensor (or small array of pixel sensors) on each detector location should suffice. What matters most for the proposed feedback loop (Fig.~\mbox{\ref{fig:screen}}) is the speed of response of the photon detectors and the photon-triggered signal chain, and the ability to minimise noise. Single-photon avalanche diode (SPAD) arrays have developed rapidly since first being demonstrated in 2003 \mbox{\cite{SPAD_origin}}. SPADs are solid-state detectors that allow photon counting with time-resolution of order tens of picoseconds \mbox{\cite{SPAD_2,SPAD_3,Charbon2014SPAD,SPAD_5,SPAD_0}}. The budget for the feedback loop is set by the total latency from a U$_4$ detection to the settled mirror or pickoff: with detector response, discrimination electronics and electro-optic switching each at the nanosecond scale, a total latency of order $100$ ns corresponds to roughly $20$ m of extra fibre on the lower arm (at $\approx 5$ ns per metre), well within the delays already demonstrated in delayed-choice experiments \mbox{\cite{Ma1221}}. Detector inefficiency and photon loss remove runs from the coincidence set rather than misclassifying them (a run in which the switch fired but no D-side detection was recorded is simply discarded), so, up to the usual fair-sampling assumption, imperfect efficiency reduces rates without biasing the measured distribution. While multiplexing to large arrays over small areas remains challenging with this detector technology, this will not be an issue for the experiment outlined in Fig.~\mbox{\ref{fig:screen}}, as large-format multiplexing will not be required.

If large-format multiplexing is found to be required for modified versions of the experiment proposed in this paper, there exist a number of state-of-the-art technologies that can achieve high-speed photon counting while providing relatively good spatial resolution. The most promising of these technologies is the microwave kinetic inductance detector ({\sc MKID}) array \mbox{\cite{arcons1, Mazin12, darkness1, strader1, Gao_thesis_2008, Eoin_2020}}. An array of {\sc MKID}s can sample the spatial domain to a resolution of better than 100 microns, while providing time resolution of around 100 ns on each pixel with zero read noise. The only false counts result from cosmic rays, but these events display a different profile and can easily be filtered. Thus, it is now feasible to simultaneously monitor a relatively large measurement plane with good spatial resolution (for a conventional quantum-eraser type experiment), as {\sc MKID}s are easily multiplexed to arrays of many thousands of pixels through frequency division multiplexing \mbox{\cite{mcHugh1, Eoin_2020}}.

The feasibility of performing a delayed-choice quantum eraser experiment with photon-counting {\sc MKID} arrays, including design considerations relevant to the setup proposed here, is discussed further in \cite{bracken2022delayed}.

The extensions of \Cref{sec:extensions} add very few additional requirements to this. \Cref{sec:delayscan}'s delay scan needs a variable fibre delay and electro-optic switching times of a few nanoseconds~\cite{Jacques2007Wheeler}, and \Cref{sec:weak}'s measurement-strength scan replaces the switchable mirror with a switchable variable-reflectivity pickoff. \Cref{sec:iterated}'s iterated photonic experiment however would be more demanding, as triplet rates in current cascaded-down-conversion sources are low \cite{Hubel2010Triplets,Hamel2014Triplets}. However, as we discuss below, we can reproduce the quantum statistics for all these experiments (if not the full physics of the scenario) using quantum circuits.

\subsection{A qubit-circuit implementation}
\label{sec:circuit}

The feedback experiment we give above is fundamentally a two-qubit protocol with mid-circuit measurement and feed-forward, so can be run on current programmable quantum hardware. Let us encode the slit label in a qubit, $\ket{{\rm A}} \to \ket{0}$ and $\ket{{\rm B}} \to \ket{1}$, so that the post-crystal state \eqref{eq:psin} is the Bell state $(\ket{00}+\ket{11})/\sqrt{2}$, the which-way measurements are computational-basis ($Z$) measurements, and the eraser measurements are $X$-basis measurements (a Hadamard followed by a $Z$ measurement). The protocol given in Fig.~\ref{fig:screen} then becomes the circuit shown in Fig.~\ref{fig:circuit}: we measure the U qubit in the $X$ basis; if the outcome is $+$ (U$_4$), then we measure the D qubit in the $Z$ basis, with the outcome $0$ playing the role of a D$_1$ click, and measure it in the $X$ basis if there is no click; if the U outcome is $-$ (U$_3$), then we measure the D qubit in the $X$ basis directly. Conditional measurement sequences of this kind are routine in dynamic circuits \cite{Corcoles2021Dynamic}, and delayed-choice experiments with causal-model analyses have already been performed on comparable platforms \cite{Yu2019CausalDelayed,Polino2019DIDelayed}.

\begin{figure}
    \centering
    \begin{tikzpicture}[line width=0.9pt]
    % wires
    \node[left] at (0,1.5) {$\ket{0}_{\rm U}$};
    \node[left] at (0,0) {$\ket{0}_{\rm D}$};
    \draw (0,1.5) -- (3.35,1.5);
    \draw (0,0) -- (4.75,0);
    % H on U
    \draw[fill=white] (0.4,1.2) rectangle (1.0,1.8); \node at (0.7,1.5) {$H$};
    % CNOT
    \filldraw (1.5,1.5) circle (0.06);
    \draw (1.5,1.5) -- (1.5,-0.18);
    \draw (1.5,0) circle (0.18);
    % H on U
    \draw[fill=white] (2.0,1.2) rectangle (2.6,1.8); \node at (2.3,1.5) {$H$};
    % meter on U
    \draw[fill=white] (3.35,1.2) rectangle (4.15,1.8);
    \draw (3.5,1.33) arc (155:25:0.26);
    \draw (3.75,1.33) -- (3.99,1.68);
    \node[above] at (3.75,1.82) {\footnotesize $-\!:$ U$_3$~~$+\!:$ U$_4$};
    % classical double wire
    \draw (4.15,1.46) -- (5.35,1.46);
    \draw (4.15,1.54) -- (5.43,1.54);
    \draw (5.35,1.46) -- (5.35,0.62);
    \draw (5.43,1.54) -- (5.43,0.62);
    % conditional measurement box on D
    \draw[dashed, fill=white] (4.75,-0.75) rectangle (7.55,0.62);
    \node[align=left] at (6.15,-0.05) {\footnotesize U$_4$: $M_Z$, then $M_X$\\[-2pt] \footnotesize \phantom{U$_4$: }if no click\\[2pt] \footnotesize U$_3$: $M_X$};
    \end{tikzpicture}
    \caption{Qubit-circuit implementation of the feedback experiment. We use a Hadamard and a {\sc CNOT} to prepare the Bell state $(\ket{00}+\ket{11})/\sqrt{2}$, encoding the path label ($\ket{{\rm A}} \to \ket{0}$, $\ket{{\rm B}} \to \ket{1}$). We then measure the U qubit in the $X$ basis (Hadamard, then a computational-basis measurement); the result is fed forward (double line) to select the measurement sequence on the D qubit: if the result is U$_4$, we perform a computational-basis measurement $M_Z$ whose $0$ outcome plays the role of a D$_1$ click, followed by an $X$-basis measurement $M_X$ if there is no click; if the result is U$_3$, we measure $M_X$. \Cref{sec:iterated}'s iterated-feedback protocol adds a third qubit, with the preparation extended to the 3-qubit GHZ state $(\ket{000}+\ket{111})/\sqrt{2}$.}
    \label{fig:circuit}
\end{figure}

Note that this quantum circuit implementation cannot enforce space-like separations or macroscopic optical delays, so it cannot fully replace the photonic experiment. It does, however, allow us to demonstrate the feedback logic and the predicted statistics (both the single-loop values from \Cref{sec:resolve} and the iterated-feedback distribution \eqref{eq:ghzdist}) on readily-available hardware.

\section{Conclusion}
\label{sec:conc}

We have here proposed a new experiment that combines the Delayed-Choice Quantum Eraser with the Elitzur-Vaidman bomb gedanken experiment and extends the combination with a feedback loop. When one tries to explain what happens in this experiment using a local realist interpretation, one is left with two choices: superdeterminism or retrocausality. The retrocausal option, taken as dynamical retrocausality, can lead to a causality paradox. When one prevents the paradox by requiring a consistent time-evolution, this form of retrocausality requires measurement outcomes which differ from the predictions of quantum mechanics. The difference is measurable and we have laid out a way to perform this measurement in the near future. Because the feedback makes the measurement setting endogenous, the experiment tests directly what a Bell test, with its external settings, cannot: whether a backwards influence of settings on earlier events acts consistently.

We have also classified the consistency-enforcing hidden-variables models of the feedback loop: they match quantum mechanics only when the backwards channel either does nothing, or is concealed by fine-tuned constraints on the hidden variables. The extensions we propose widen the test in switching delay, in measurement strength, and in feedback depth. Further, the logic of our protocol, and its extensions, can be checked on current quantum hardware using a qubit-circuit implementation.

\textit{Acknowledgements:}
We thank Tim Palmer and John Rarity for useful comments.
CB acknowledges support by Enterprise Ireland under the HEU award, grant number EI/CS20212057-BRACKEN. JRH acknowledges support from the University of York's EPSRC DTP grant EP/R513386/1, the Quantum Communications Hub funded by the EPSRC grants EP/M013472/1 and EP/T001011/1, Hiroshima University's Phoenix Postdoctoral Fellowship for Research programme, a Royal Society Research Grant (RG/R1/251590), an EPSRC Mathematical Sciences Small Grant (UKRI3647), and from their EPSRC Quantum Technologies Career Acceleration Fellowship (UKRI1217).
SH acknowledges support by the Deutsche Forschungsgemeinschaft (DFG, German Research Foundation) under grant number HO 2601/8-1.

\textit{Competing Interest Statement:} The authors declare that there are no competing interests.

\textit{Code and Data Availability:} A Mathematica notebook verifying the quantitative claims made here is available on Zenodo~\cite{QuantumEraserCode}.

\textit{Author Contribution Statement:} All authors contributed to this article equally. Authors are listed alphabetically.

\bibliographystyle{unsrturl}
\bibliography{ref}

@article{adlam2022two,
  title={Two roads to retrocausality},
  author={Adlam, Emily},
  journal={Synthese},
  volume={200},
  number={5},
  pages={422},
  year={2022},
  publisher={Springer},
  doi={10.1007/s11229-022-03919-0}
}

@article{Adlam2023Taxonomy,
    title={Taxonomy for Physics Beyond Quantum Mechanics},
    author={Adlam, Emily and Hance, Jonte R. and Hossenfelder, Sabine and Palmer, Tim N.},
    journal={Proceedings of the Royal Society A},
    year={2024},
    volume = {480},
    number = {2294},
    pages = {20230779},
    doi={10.1098/rspa.2023.0779}
}

@article{Almada2015Retrocausal,
  title={Are retrocausal accounts of entanglement unnaturally fine-tuned?},
  author={Almada, D and Ch'ng, K and Kintner, S and Morrison, B and Wharton, KB},
  journal={arXiv preprint arXiv:1510.03706},
  year={2015},
  url={https://arxiv.org/abs/1510.03706}
}

@article{Arntzenius1994-ARNSC,
	title = {Spacelike Connections},
	pages = {201--217},
	journal = {British Journal for the Philosophy of Science},
	author = {Frank Arntzenius},
	publisher = {Taylor \& Francis},
	year = {1994},
	doi = {10.1093/bjps/45.1.201},
	volume = {45},
	number = {1}
}

@article{Aspect_1,
  title = {Experimental Realization of {E}instein-{P}odolsky-{R}osen-{B}ohm Gedankenexperiment: A New Violation of {B}ell's Inequalities},
  author = {Aspect, Alain and Grangier, Philippe and Roger, G\'erard},
  journal = {Phys. Rev. Lett.},
  volume = {49},
  issue = {2},
  pages = {91--94},
  numpages = {0},
  year = {1982},
  month = {Jul},
  publisher = {American Physical Society},
  doi = {10.1103/PhysRevLett.49.91}
}

@inproceedings{Eoin_2020,
author = {Baldwin, Eoin and De Lucia, Mario and Bracken, Colm and Ulbricht, Gerhard and Ray, Tom},
title = {{Frequency domain multiplexing with the Xilinx ZCU111 RFSoC board}},
volume = {11454},
booktitle = {X-Ray, Optical, and Infrared Detectors for Astronomy IX},
organization = {International Society for Optics and Photonics},
publisher = {SPIE},
year = {2020},
doi = {10.1117/12.2561108}
}

@article{Bell1964OnEPR,
  title={On the {E}instein {P}odolsky {R}osen paradox},
  author={Bell, John S},
  journal={Physics Physique Fizika},
  volume={1},
  number={3},
  pages={195},
  year={1964},
  publisher={APS},
  doi={10.1103/PhysicsPhysiqueFizika.1.195}
}

@book{Bell2004Speakable,
  title={Speakable and unspeakable in quantum mechanics: Collected papers on quantum philosophy},
  author={Bell, John S},
  year={2004},
  publisher={Cambridge University Press},
  edition={2}
}

@article{ben2007impossibility,
  title={The impossibility of backwards causation},
  author={Ben-Yami, Hanoch},
  journal={The Philosophical Quarterly},
  volume={57},
  number={228},
  pages={439--455},
  year={2007},
  publisher={Blackwell Publishers Ltd Oxford, UK and Boston, USA},
  doi={10.1111/j.1467-9213.2007.493.x}
}

@incollection{berkovitz2011explanation,
  title={On explanation in retro-causal interpretations of quantum mechanics},
  author={Berkovitz, Joseph},
  booktitle={Probabilities, Causes and Propensities in Physics},
  pages={115--155},
  year={2011},
  publisher={Springer},
  doi={10.1007/978-1-4020-9904-5_6}
}

@article{bracken2022delayed,
  title={A Delayed-Choice Quantum Eraser with Photon-Counting MKIDs (Experimental Design)},
  author={Bracken, CP and McAleer, C},
  journal={Journal of Low Temperature Physics},
  volume={209},
  number={5-6},
  pages={899--911},
  year={2022},
  publisher={Springer},
  doi={10.1007/s10909-022-02852-3}
}

@misc{QuantumEraserCode,
    title = {Code for ``The Quantum Eraser Paradox''},
    author = {Bracken, Colm and Hance, Jonte R. and Hossenfelder, S},
    year = {2026},
    howpublished = {Zenodo},
    doi={10.5281/zenodo.23159117}
}

@article{SPAD_0,
author = {Bruschini, Claudio and Homulle, Harald and Antolovic, Ivan Michel and Burri, Samuel and Charbon, Edoardo},
title = {Single-photon avalanche diode imagers in biophotonics: review and outlook},
journal = {Light: Science \& Applications},
  volume = {8},
  issue = {1},  
  year = {2019},
  publisher = {Nature},
  doi = {10.1038/s41377-019-0191-5}
}

@article{Catani2021Interference,
doi = {10.22331/q-2023-09-25-1119},
  title = {Why interference phenomena do not capture the essence of quantum theory},
  author = {Catani, Lorenzo and Leifer, Matthew and Schmid, David and Spekkens, Robert W.},
  journal = {{Quantum}},
  issn = {2521-327X},
  publisher = {{Verein zur F{\"{o}}rderung des Open Access Publizierens in den Quantenwissenschaften}},
  volume = {7},
  pages = {1119},
  month = sep,
  year = {2023}
}

@article{catani2022reply,
  title={Reply to ``Comment on `Why interference phenomena do not capture the essence of quantum theory'{}''},
  author={Catani, Lorenzo and Leifer, Matthew and Schmid, David and Spekkens, Robert W},
  journal={arXiv preprint arXiv:2207.11791},
  year={2022},
  doi={10.48550/arXiv.2207.11791}
}

@article{Charbon2014SPAD,
  title={Single-photon imaging in complementary metal oxide semiconductor processes},
  author={Charbon, E},
  journal={Philosophical Transactions of the Royal Society A: Mathematical, Physical and Engineering Sciences},
  volume={372},
  number={2012},
  pages={20130100},
  year={2014},
  publisher={The Royal Society Publishing},
  doi={10.1098/rsta.2013.0100}
}

@article{Chaves2018CausalDC,
  title = {Causal Modeling the Delayed-Choice Experiment},
  author = {Chaves, Rafael and Lemos, Gabriela Barreto and Pienaar, Jacques},
  journal = {Phys. Rev. Lett.},
  volume = {120},
  issue = {19},
  pages = {190401},
  numpages = {6},
  year = {2018},
  month = {May},
  publisher = {American Physical Society},
  doi = {10.1103/PhysRevLett.120.190401}
}

@article{Cohen2020RetrocausalI,
  title = {Realism and causality. I. Pilot wave and retrocausal models as possible facilitators},
  author = {Cohen, Eliahu and Cort\^es, Marina and Elitzur, Avshalom and Smolin, Lee},
  journal = {Phys. Rev. D},
  volume = {102},
  issue = {12},
  pages = {124027},
  numpages = {16},
  year = {2020},
  month = {Dec},
  publisher = {American Physical Society},
  doi = {10.1103/PhysRevD.102.124027}
}

@article{Cohen2020RetrocausalII,
  title = {Realism and causality. II. Retrocausality in energetic causal sets},
  author = {Cohen, Eliahu and Cort\^es, Marina and Elitzur, Avshalom C. and Smolin, Lee},
  journal = {Phys. Rev. D},
  volume = {102},
  issue = {12},
  pages = {124028},
  numpages = {19},
  year = {2020},
  month = {Dec},
  publisher = {American Physical Society},
  doi = {10.1103/PhysRevD.102.124028}
}

@article{Corry2015Retrocausal,
  title={Retrocausal models for EPR},
  author={Corry, Richard},
  journal={Studies in History and Philosophy of Science Part B: Studies in History and Philosophy of Modern Physics},
  volume={49},
  pages={1--9},
  year={2015},
  publisher={Elsevier},
  doi={10.1016/j.shpsb.2014.11.001}
}

@article{Costa1977TimeI,
  title={Time symmetry and the {E}instein paradox},
  author={de Beauregard, O Costa},
  journal={Il Nuovo Cimento B (1971-1996)},
  volume={42},
  number={1},
  pages={41--64},
  year={1977},
  publisher={Springer},
  doi={10.1007/BF02906749}
}

@article{Costa1979TimeII,
  title={Time symmetry and the {E}instein paradox.---{II}},
  author={de Beauregard, O Costa},
  journal={Il Nuovo Cimento B (1971-1996)},
  volume={51},
  number={2},
  pages={267--279},
  year={1979},
  publisher={Springer},
  doi={10.1007/bf02743436}
}

@article{Costa1985Concerning,
  title={On some frequent but controversial statements concerning the {E}instein-{P}odolsky-{R}osen correlations},
  author={de Beauregard, O Costa},
  journal={Foundations of Physics},
  volume={15},
  number={8},
  pages={871--887},
  year={1985},
  publisher={Springer},
  doi={10.1007/BF00738320}
}

@article{Costa1987Zigzagging,
  title={On the zigzagging causality EPR model: Answer to Vigier and coworkers and to Sutherland},
  author={de Beauregard, O Costa},
  journal={Foundations of physics},
  volume={17},
  number={8},
  pages={775--785},
  year={1987},
  publisher={Springer},
  doi={10.1007/BF00733266}
}

@Article{Dieguez2022Realism,
author={Dieguez, Pedro R. and Guimar{\~a}es, J{\'e}ferson R. and Peterson, John P. S. and Angelo, Renato M. and Serra, Roberto M.},
title={Experimental assessment of physical realism in a quantum-controlled device},
journal={Communications Physics},
year={2022},
month={Apr},
day={06},
volume={5},
number={1},
pages={82},
issn={2399-3650},
doi={10.1038/s42005-022-00828-z}
}

@article{Dowe1992Asymmetry,
  title={Process causality and asymmetry},
  author={Dowe, Phil},
  journal={Erkenntnis},
  volume={37},
  number={2},
  pages={179--196},
  year={1992},
  publisher={Springer},
  doi={10.1007/BF00209321}
}

@article{Dowe1996Backwards,
  title={Backwards causation and the direction of causal processes},
  author={Dowe, Phil},
  journal={Mind},
  volume={105},
  number={418},
  pages={227--248},
  year={1996},
  publisher={Oxford University Press},
  doi={10.1093/mind/105.418.227}
}

@article{Dowe1997Defense,
  title={A defense of backwards in time causation models in quantum mechanics},
  author={Dowe, Phil},
  journal={Synthese},
  volume={112},
  number={2},
  pages={233--246},
  year={1997},
  publisher={Springer},
  doi={10.1023/A:1004932911141}
}

@article{BBO_1,
author = {Eimerl,D.  and Davis,L.  and Velsko,S.  and Graham,E. K.  and Zalkin,A. },
title = {Optical, mechanical, and thermal properties of barium borate},
journal = {Journal of Applied Physics},
volume = {62},
number = {5},
pages = {1968-1983},
year = {1987},
doi = {10.1063/1.339536}
}

@Article{Elitzur1993Bomb,
author="Elitzur, Avshalom C.
and Vaidman, Lev",
title="Quantum mechanical interaction-free measurements",
journal="Foundations of Physics",
year="1993",
month="Jul",
day="01",
volume="23",
number="7",
pages="987--997",
issn="1572-9516",
doi="10.1007/BF00736012"
}

@article{ellerman2015delayed,
  title={Why delayed choice experiments do not imply retrocausality},
  author={Ellerman, David},
  journal={Quantum Studies: Mathematics and Foundations},
  volume={2},
  number={2},
  pages={183--199},
  year={2015},
  publisher={Springer},
  doi={10.1007/s40509-014-0026-2}
}

@article{Evans2015Retrocausality,
  title={Retrocausality at no extra cost},
  author={Evans, Peter W},
  journal={Synthese},
  volume={192},
  number={4},
  pages={1139--1155},
  year={2015},
  publisher={Springer},
  doi={10.1007/s11229-014-0605-0}
}

@phdthesis{Gao_thesis_2008,
  author        = "Gao, Jiansong",
  title         = "The Physics of Superconducting Microwave Resonators",
  school        = "California Institute of Technology",
  address       = "Califoria, U.S.A.",
  month         = May,
  year          = "2008",
  doi = "10.7907/RAT0-VM75"
}

@article{GiustinaLoophole2015,
  title = {Significant-Loophole-Free Test of {B}ell's Theorem with Entangled Photons},
  author = {Giustina, Marissa and Versteegh, Marijn A. M. and Wengerowsky, S\"oren and Handsteiner, Johannes and Hochrainer, Armin and Phelan, Kevin and Steinlechner, Fabian and Kofler, Johannes and Larsson, Jan-\AA{}ke and Abell\'an, Carlos and Amaya, Waldimar and Pruneri, Valerio and Mitchell, Morgan W. and Beyer, J\"orn and Gerrits, Thomas and Lita, Adriana E. and Shalm, Lynden K. and Nam, Sae Woo and Scheidl, Thomas and Ursin, Rupert and Wittmann, Bernhard and Zeilinger, Anton},
  journal = {Phys. Rev. Lett.},
  volume = {115},
  issue = {25},
  pages = {250401},
  numpages = {7},
  year = {2015},
  month = {Dec},
  publisher = {American Physical Society},
  doi = {10.1103/PhysRevLett.115.250401}
}

@article{Hance2021Ensemble,
  title={The wave function as a true ensemble},
  author={Hance, Jonte R. and Hossenfelder, Sabine},
  journal={Proceedings of the Royal Society A},
  volume={478},
  pages={20210705},
  year={2022},
  publisher={Royal Society},
  doi={10.1098/rspa.2021.0705}
}

@article{Hance2021Quantum,
  title={How Quantum is Quantum Counterfactual Communication?},
  author={Hance, Jonte R and Ladyman, James and Rarity, John},
  journal={Foundations of Physics},
  year={2021},
  doi={10.1007/s10701-021-00412-5},
  volume={51},
  number={1},
  pages={12},
  month={Feb}
}

@article{Hance2021CFGI,
  title={Counterfactual ghost imaging},
  author={Hance, Jonte R and Rarity, John},
  journal={npj Quantum Information},
  volume={7},
  number={1},
  pages={1--7},
  year={2021},
  publisher={Nature Publishing Group},
  doi={10.1038/s41534-021-00411-4}
}

@Article{Hance2022Supermeasured,
author={Hance, Jonte R. and Hossenfelder, Sabine and Palmer, Tim N.},
title={Supermeasured: Violating {B}ell-Statistical Independence Without Violating Physical Statistical Independence},
journal={Foundations of Physics},
year={2022},
month={Jul},
day={19},
volume={52},
number={4},
pages={81},
issn={1572-9516},
doi={10.1007/s10701-022-00602-9}
}

@article{hance2022comment,
  title={Comment on ``Why interference phenomena do not capture the essence of quantum theory'' by {C}atani et al.},
  author={Hance, Jonte R and Hossenfelder, Sabine},
  journal={arXiv preprint arXiv:2204.01768},
  year={2022},
  doi={10.48550/arXiv.2204.01768}
}

@misc{hance2026noncontextualversuscontextualinterferometry,
      title={Noncontextual versus contextual interferometry}, 
      author={Jonte R. Hance and Jakov Krnic and Jan-Åke Larsson},
      year={2026},
      eprint={2601.13109},
      archivePrefix={arXiv},
      primaryClass={quant-ph},
      url={https://arxiv.org/abs/2601.13109}, 
}

@article{Hensen2015loophole,
  title={Loophole-free {B}ell inequality violation using electron spins separated by 1.3 kilometres},
  author={Hensen, Bas and Bernien, Hannes and Dr{\'e}au, Ana{\"\i}s E and Reiserer, Andreas and Kalb, Norbert and Blok, Machiel S and Ruitenberg, Just and Vermeulen, Raymond FL and Schouten, Raymond N and Abell{\'a}n, Carlos and others},
  journal={Nature},
  volume={526},
  number={7575},
  pages={682--686},
  year={2015},
  publisher={Nature Publishing Group},
  doi={10.1038/nature15759}
}

@article{Hossenfelder2020Rethinking,
  title={Rethinking superdeterminism},
  author={Hossenfelder, Sabine and Palmer, Tim},
  journal={Frontiers in Physics},
  volume={8},
  pages={139},
  year={2020},
  publisher={Frontiers},
  doi={10.3389/fphy.2020.00139}
}

@article{Hossenfelder2020Perplexed,
  title={Superdeterminism: A Guide for the Perplexed},
  author={Hossenfelder, Sabine},
  journal={arXiv preprint arXiv:2010.01324},
  year={2020},
  url={https://arxiv.org/abs/2010.01324}
}

@article{Hosten2006CFComp,
  title={Counterfactual quantum computation through quantum interrogation},
  author={Hosten, Onur and Rakher, Matthew T and Barreiro, Julio T and Peters, Nicholas A and Kwiat, Paul G},
  journal={Nature},
  volume={439},
  number={7079},
  pages={949--952},
  year={2006},
  publisher={Nature Publishing Group},
  doi={10.1038/nature04523}
}

@article{Karab2018SPDC,
author = {Karan, Suman and Aarav, Shaurya and Bharadhwaj, Homanga and Taneja, Lavanya and De, Arinjoy and Kulkarni, Girish and Meher, Nilakantha and Jha, Anand K.},
year = {2020},
title = {Phase matching in $\beta$-barium borate crystals for spontaneous parametric down-conversion},
journal = {Journal of Optics},
volume = {22},
number = {8},
pages = {083501},
publisher = {IOP Publishing},
doi = {10.1088/2040-8986/ab89e4}
}

@Article{Kastner2019,
author={Kastner, R. E.},
title={The `Delayed Choice Quantum Eraser' Neither Erases Nor Delays},
journal={Foundations of Physics},
year={2019},
month={Jul},
day={01},
volume={49},
number={7},
pages={717-727},
issn={1572-9516},
doi={10.1007/s10701-019-00278-8},
url={https://doi.org/10.1007/s10701-019-00278-8}
}

@article{Kim2000Delayed,
  title={Delayed ``choice'' quantum eraser},
  author={Kim, Yoon-Ho and Yu, Rong and Kulik, Sergei P and Shih, Yanhua and Scully, Marlan O},
  journal={Physical Review Letters},
  volume={84},
  number={1},
  pages={1},
  year={2000},
  publisher={APS},
  doi={10.1103/PhysRevLett.84.1}
}

@article{down_conversion_1,
  title = {New High-Intensity Source of Polarization-Entangled Photon Pairs},
  author = {Kwiat, Paul G. and Mattle, Klaus and Weinfurter, Harald and Zeilinger, Anton and Sergienko, Alexander V. and Shih, Yanhua},
  journal = {Phys. Rev. Lett.},
  volume = {75},
  issue = {24},
  pages = {4337--4341},
  numpages = {0},
  year = {1995},
  month = {Dec},
  publisher = {American Physical Society},
  doi = {10.1103/PhysRevLett.75.4337}
}

@article {Ma1221,
	author = {Ma, Xiao-Song and Kofler, Johannes and Qarry, Angie and Tetik, Nuray and Scheidl, Thomas and Ursin, Rupert and Ramelow, Sven and Herbst, Thomas and Ratschbacher, Lothar and Fedrizzi, Alessandro and Jennewein, Thomas and Zeilinger, Anton},
	title = {Quantum erasure with causally disconnected choice},
	volume = {110},
	number = {4},
	pages = {1221--1226},
	year = {2013},
	doi = {10.1073/pnas.1213201110},
	publisher = {National Academy of Sciences},
	issn = {0027-8424},
	journal = {Proceedings of the National Academy of Sciences}
}

@book{maudlin2011quantum,
  title={Quantum non-locality and relativity: Metaphysical intimations of modern physics},
  author={Maudlin, Tim},
  year={2011},
  publisher={John Wiley \& Sons},
  isbn={978-1-444-39696-6}
}

@INPROCEEDINGS{arcons1,
   author = {{Mazin}, B.~A. and {O'Brien}, K. and {McHugh}, S. and {Bumble}, B. and 
	{Moore}, D. and {Golwala}, S. and {Zmuidzinas}, J.},
    title = "{ARCONS: a highly multiplexed superconducting optical to near-IR camera}",
booktitle = {Ground-based and Airborne Instrumentation for Astronomy III},
     year = {2010},
   series = {Proc. of SPIE},
   volume = {7735},
archivePrefix = "arXiv",
   eprint = {1007.0752},
 primaryClass = "astro-ph.IM",
    month = jul,
      eid = {773518},
    pages = {773518},
      doi = {10.1117/12.856440}
}

@article{Mazin12,
author = {Benjamin A. Mazin and Bruce Bumble and Seth R. Meeker and Kieran O'Brien and Sean McHugh and Eric Langman},
journal = {Opt. Express},
number = {2},
pages = {1503--1511},
publisher = {OSA},
title = {A superconducting focal plane array for ultraviolet, optical, and near-infrared astrophysics},
volume = {20},
month = {Jan},
year = {2012},
doi = {10.1364/OE.20.001503},
}

@article{mcHugh1,
  author={McHugh, Sean and Mazin, Benjamin A and Serfass, Bruno and Meeker, Seth and O’Brien, Kieran and Duan, Ran and Raffanti, Rick and Werthimer, Dan},
  title         = "A readout for large arrays of microwave kinetic inductance detectors",
  journal       = "Review of Scientific Instruments",
  volume        = "83",
  issue         = "4",
  pages         ={044702},
  year          = "2012",
  doi           = "10.1063/1.3700812",
  publisher={American Institute of Physics}
}

@article{darkness1,
   author = "Meeker,Seth and others",
    title = "{DARKNESS: A Microwave Kinetic Inductance Detector Integral Field Spectrograph for High-contrast Astronomy}",
  journal = {Publications of the Astronomical Society of the Pacific},
     year = 2018,
    month = jun,
   volume = 130,
   number = 988,
    pages = {065001},
      doi = {10.1088/1538-3873/aab5e7},
}

@article{Norsen2011Bell,
  title={{J}ohn {S}. {B}ell's concept of local causality},
  author={Norsen, Travis},
  journal={American Journal of Physics},
  volume={79},
  number={12},
  pages={1261--1275},
  year={2011},
  publisher={American Association of Physics Teachers},
  doi={10.1119/1.3630940}
}

@book{Pearl2009Causality,
  title={Causality},
  author={Pearl, Judea},
  year={2009},
  publisher={Cambridge university press},
  isbn={052189560X}
}

@article{SPAD_5,
  title={Compact SPAD-based pixel architectures for time-resolved image sensors},
  author={Perenzoni, Matteo and Pancheri, Lucio and Stoppa, David},
  journal={Sensors},
  volume={16},
  number={5},
  pages={745},
  year={2016},
  publisher={Multidisciplinary Digital Publishing Institute},
  doi={10.3390/s16050745}
}

@article{Price1984AffectingPast,
  title={The philosophy and physics of affecting the past},
  author={Price, Huw},
  journal={Synthese},
  volume={61},
  pages={299--323},
  year={1984},
  publisher={Springer},
  doi={10.1007/BF00485056}
}

@article{Price1994Neglected,
  title={A neglected route to realism about quantum mechanics},
  author={Price, Huw},
  journal={Mind},
  volume={103},
  number={411},
  pages={303--336},
  year={1994},
  publisher={Oxford University Press},
  doi={10.1093/mind/103.411.303}
}

@book{Price1996Time,
  title={Time's arrow \& Archimedes' point: new directions for the physics of time},
  author={Price, Huw},
  year={1996},
  publisher={Oxford University Press},
  isbn={0199839328}
}

@article{Price2008ToyModel,
title = {Toy models for retrocausality},
journal = {Studies in History and Philosophy of Science Part B: Studies in History and Philosophy of Modern Physics},
volume = {39},
number = {4},
pages = {752-761},
year = {2008},
issn = {1355-2198},
doi = {10.1016/j.shpsb.2008.05.006},
author = {Price, Huw}
}

@article{Price2012SymmImpliesRetrocausality,
  title={Does time-symmetry imply retrocausality? How the quantum world says ``Maybe''?},
  author={Price, Huw},
  journal={Studies in History and Philosophy of Science Part B: Studies in History and Philosophy of Modern Physics},
  volume={43},
  number={2},
  pages={75--83},
  year={2012},
  publisher={Elsevier},
  doi={10.1016/j.shpsb.2011.12.003}
}

@Article{Price2015Disentangling,
AUTHOR = {Price, Huw and Wharton, Ken},
TITLE = {Disentangling the Quantum World},
JOURNAL = {Entropy},
VOLUME = {17},
YEAR = {2015},
NUMBER = {11},
PAGES = {7752--7767},
DOI = {10.3390/e17117752}
}

@article{Qureshi2020Demystifying,
  title={Demystifying the delayed-choice quantum eraser},
  author={Qureshi, Tabish},
  journal={European Journal of Physics},
  volume={41},
  number={5},
  pages={055403},
  year={2020},
  publisher={IOP Publishing},
  doi={10.1088/1361-6404/ab923e}
}

@article{Qureshi2021Delayed,
  title={The delayed-choice quantum eraser leaves no choice},
  author={Qureshi, Tabish},
  journal={International Journal of Theoretical Physics},
  volume={60},
  number={8},
  pages={3076--3086},
  year={2021},
  publisher={Springer},
  doi={10.1007/s10773-021-04906-w}
}

@article{rivlin2022quantum,
  title={The Quantum Eraser from a Weak Values Perspective},
  author={Rivlin, Tom},
  journal={arXiv preprint arXiv:2201.10320},
  year={2022},
  doi={10.48550/arXiv.2201.10320}
}

@article{SPAD_origin,
  title={Single photon detector fabricated in a complementary metal--oxide--semiconductor high-voltage technology},
  author={Rochas, A and Gani, M and Furrer, B and Besse, PA and Popovic, RS and Ribordy, G and Gisin, N},
  journal={Review of Scientific Instruments},
  volume={74},
  number={7},
  pages={3263--3270},
  year={2003},
  publisher={American Institute of Physics},
  doi={10.1063/1.1584083}
}

@article{Salih2013Protocol,
  title = {Protocol for Direct Counterfactual Quantum Communication},
  author = {Salih, Hatim and Li, Zheng-Hong and Al-Amri, Mohammad and Zubairy, Muhammad Suhail},
  journal = {Phys. Rev. Lett.},
  volume = {110},
  issue = {17},
  pages = {170502},
  numpages = {5},
  year = {2013},
  month = {Apr},
  publisher = {American Physical Society},
  doi = {10.1103/PhysRevLett.110.170502}
}

@article{Salih2016Qubit,
  title={Protocol for counterfactually transporting an unknown qubit},
  author={Salih, Hatim},
  journal={Frontiers in Physics},
  volume={3},
  pages={94},
  year={2016},
  publisher={Frontiers},
  doi={10.3389/fphy.2015.00094}
}

@article{Salih2018Laws,
    author={Salih, Hatim and McCutcheon, Will and Hance, Jonte R. and Rarity, John},
title={The laws of physics do not prohibit counterfactual communication},
journal={npj Quantum Information},
year={2022},
month={May},
day={18},
volume={8},
number={1},
pages={60},
issn={2056-6387},
doi={10.1038/s41534-022-00564-w}
}

@article{Salih2021EFQubit,
	doi = {10.1088/1367-2630/abd3c4},
	url = {https://doi.org/10.1088/1367-2630/abd3c4},
	year = 2021,
	month = {jan},
	publisher = {{IOP} Publishing},
	volume = {23},
	number = {1},
	pages = {013004},
	author = {Hatim Salih and Jonte R Hance and Will McCutcheon and Terry Rudolph and John Rarity},
	title = {Exchange-free computation on an unknown qubit at a distance},
	journal = {New Journal of Physics}
}

@article{Salih2020DetTele,
      title={Deterministic Teleportation and Universal Computation Without Particle Exchange}, 
      author={Hatim Salih and Jonte R. Hance and Will McCutcheon and Terry Rudolph and John Rarity},
      year={2020},
      journal={arXiv preprint arXiv:2009.05564},
      url={https://arxiv.org/abs/2009.05564}
}

@article{Scarcelli_2007,
   title={Random delayed-choice quantum eraser via two-photon imaging},
   volume={44},
   ISSN={1434-6079},
   DOI={10.1140/epjd/e2007-00164-y},
   number={1},
   journal={The European Physical Journal D},
   publisher={Springer Science and Business Media LLC},
   author={Scarcelli, G. and Zhou, Y. and Shih, Y.},
   year={2007},
   month={May},
   pages={167–173}
}

@article{Scully1982Eraser,
  title={Quantum eraser: A proposed photon correlation experiment concerning observation and ``delayed choice'' in quantum mechanics},
  author={Scully, Marlan O and Dr{\"u}hl, Kai},
  journal={Physical Review A},
  volume={25},
  number={4},
  pages={2208},
  year={1982},
  publisher={APS},
  doi={10.1103/PhysRevA.25.2208}
}

@article{Schmid2020Unscrambling,
  title={Unscrambling the omelette of causation and inference: The framework of causal-inferential theories},
  author={Schmid, David and Selby, John H and Spekkens, Robert W},
  journal={arXiv preprint arXiv:2009.03297},
  year={2020},
  url={https://arxiv.org/abs/2009.03297}
}

@article{Sen2019Retrocausal,
  title={A Local $\psi$-Epistemic Retrocausal Hidden-Variable Model of {B}ell Correlations with Wavefunctions in Physical Space},
  author={Sen, Indrajit},
  journal={Foundations of Physics},
  volume={49},
  number={2},
  pages={83--95},
  year={2019},
  publisher={Springer},
  doi={10.1007/s10701-018-0231-7}
}

@inbook{Shimony1993NoSig,
place={Cambridge},
title={Controllable and uncontrollable non-locality},
volume={2},
DOI={10.1017/CBO9781139172196.010},
booktitle={The Search for a Naturalistic World View},
publisher={Cambridge University Press},
author={Shimony, Abner},
year={1993},
pages={130–139},
isbn={0521377447}
}

@book{Spirtes2000Causation,
  title={Causation, prediction, and search},
  author={Spirtes, Peter and Glymour, Clark N and Scheines, Richard and Heckerman, David},
  year={2000},
  publisher={MIT press},
  isbn={0262194406}
}

@phdthesis{strader1,
  author        = "M. Strader",
  title         = "Digital Readout for Microwave Kinetic Inductance Detectors and Applications in High Time Resolution Astronomy",
  school        = "University of California Santa Barbara",
  address       = "Califoria, U.S.A.",
  month         = "September",
  year          = "2016",
  isbn={978-1-369-34052-5}
}

@article{Sutherland1983BackwardsinTime,
  title={{B}ell's theorem and backwards-in-time causality},
  author={Sutherland, Roderick I},
  journal={International Journal of Theoretical Physics},
  volume={22},
  number={4},
  pages={377--384},
  year={1983},
  publisher={Springer},
  doi={10.1007/BF02082904}
}

@article{Sutherland1998Density,
  title={Density formalism for quantum theory},
  author={Sutherland, Roderick I},
  journal={Foundations of Physics},
  volume={28},
  number={7},
  pages={1157--1190},
  year={1998},
  publisher={Springer},
  doi={10.1023/A:1018850120826}
}

@article{Sutherland2017HowRetroHelps,
author = {Sutherland,Roderick I. },
title = {How retrocausality helps},
journal = {AIP Conference Proceedings},
volume = {1841},
number = {1},
pages = {020001},
year = {2017},
doi = {10.1063/1.4982765}
}

@article{Walborn_2002,
   title={Double-slit quantum eraser},
   volume={65},
   ISSN={1094-1622},
   DOI={10.1103/physreva.65.033818},
   number={3},
   journal={Physical Review A},
   publisher={American Physical Society (APS)},
   author={Walborn, S. P. and Terra Cunha, M. O. and Pádua, S. and Monken, C. H.},
   pages={033818},
   year={2002},
   month={Feb}
}

@article{Wharton2018NewRetro,
  title={A new class of retrocausal models},
  author={Wharton, Ken},
  journal={Entropy},
  volume={20},
  number={6},
  pages={410},
  year={2018},
  publisher={Multidisciplinary Digital Publishing Institute},
  doi={10.3390/e20060410}
}

@incollection{wheeler1978past,
  title={The “past” and the “delayed-choice” double-slit experiment},
  author={Wheeler, John Archibald},
  booktitle={Mathematical foundations of quantum theory},
  pages={9--48},
  year={1978},
  publisher={Elsevier},
  isbn={0323141188}
}

@article{Wood2015FineTuning,
  title={The lesson of causal discovery algorithms for quantum correlations: Causal explanations of {B}ell-inequality violations require fine-tuning},
  author={Wood, Christopher J and Spekkens, Robert W},
  journal={New Journal of Physics},
  volume={17},
  number={3},
  pages={033002},
  year={2015},
  publisher={IOP Publishing},
  doi={10.1088/1367-2630/17/3/033002}
}

@article{SPAD_2,
  title={Principles and features of single-photon avalanche diode arrays},
  author={Zappa, Franco and Tisa, Simone and Tosi, Alberto and Cova, Sergio},
  journal={Sensors and Actuators A: Physical},
  volume={140},
  number={1},
  pages={103--112},
  year={2007},
  publisher={Elsevier},
  doi={10.1016/j.sna.2007.06.021}
}

@inproceedings{SPAD_3,
author = {F. Zappa and A. Tosi and A. Dalla Mora and F. Guerrieri and S. Tisa},
title = {{Single-photon avalanche diode arrays and CMOS microelectronics for counting, timing, and imaging quantum events}},
volume = {7608},
booktitle = {Quantum Sensing and Nanophotonic Devices VII},
editor = {Manijeh Razeghi and Rengarajan Sudharsanan and Gail J. Brown},
organization = {International Society for Optics and Photonics},
publisher = {SPIE},
pages = {717 -- 731},
year = {2010},
doi = {10.1117/12.840362}
}

@article{Ma2016DelayedReview,
  title = {Delayed-choice gedanken experiments and their realizations},
  author = {Ma, Xiao-Song and Kofler, Johannes and Zeilinger, Anton},
  journal = {Rev. Mod. Phys.},
  volume = {88},
  issue = {1},
  pages = {015005},
  year = {2016},
  month = {Mar},
  publisher = {American Physical Society},
  doi = {10.1103/RevModPhys.88.015005}
}

@article{Fankhauser2019Taming,
  title={Taming the Delayed Choice Quantum Eraser},
  author={Fankhauser, Johannes},
  journal={Quanta},
  volume={8},
  number={1},
  pages={44--56},
  year={2019},
  doi={10.12743/quanta.v8i1.88}
}

@article{LaCour2021Classical,
  title = {Classical model of a delayed-choice quantum eraser},
  author = {La Cour, Brian R. and Yudichak, Thomas W.},
  journal = {Phys. Rev. A},
  volume = {103},
  issue = {6},
  pages = {062213},
  year = {2021},
  month = {Jun},
  publisher = {American Physical Society},
  doi = {10.1103/PhysRevA.103.062213}
}

@article{Polino2019DIDelayed,
  title = {Device-independent test of a delayed choice experiment},
  author = {Polino, Emanuele and Agresti, Iris and Poderini, Davide and Carvacho, Gonzalo and Milani, Giorgio and Lemos, Gabriela Barreto and Chaves, Rafael and Sciarrino, Fabio},
  journal = {Phys. Rev. A},
  volume = {100},
  issue = {2},
  pages = {022111},
  year = {2019},
  month = {Aug},
  publisher = {American Physical Society},
  doi = {10.1103/PhysRevA.100.022111}
}

@article{Yu2019CausalDelayed,
  title = {Realization of a causal-modeled delayed-choice experiment using single photons},
  author = {Yu, Shang and Sun, Yong-Nan and Liu, Wei and Liu, Zhao-Di and Ke, Zhi-Jin and Wang, Yi-Tao and Tang, Jian-Shun and Li, Chuan-Feng and Guo, Guang-Can},
  journal = {Phys. Rev. A},
  volume = {100},
  issue = {1},
  pages = {012115},
  year = {2019},
  month = {Jul},
  publisher = {American Physical Society},
  doi = {10.1103/PhysRevA.100.012115}
}

@article{Drezet2021NonParadox,
  title={The Quantum Eraser Non-Paradox: a comment on arXiv:2111.09347},
  author={Drezet, Aur\'elien},
  journal={arXiv preprint arXiv:2111.13357},
  year={2021},
  doi={10.48550/arXiv.2111.13357}
}

@article{Bracken2021Reply,
  title={Reply to arXiv:2111.13357 (``The Quantum Eraser Non-Paradox'')},
  author={Bracken, Colm and Hance, Jonte R. and Hossenfelder, Sabine},
  journal={arXiv preprint arXiv:2112.00436},
  year={2021},
  doi={10.48550/arXiv.2112.00436}
}

@article{Ridley2025TimeEvent,
  title={Time and event symmetry in quantum mechanics},
  author={Ridley, Michael and Adlam, Emily},
  journal={Quantum Studies: Mathematics and Foundations},
  volume={12},
  number={1},
  pages={9},
  year={2025},
  publisher={Springer},
  doi={10.1007/s40509-024-00358-z}
}

@article{Leifer2017TimeSymmetric,
  title={Is a time symmetric interpretation of quantum theory possible without retrocausality?},
  author={Leifer, Matthew S. and Pusey, Matthew F.},
  journal={Proceedings of the Royal Society A},
  volume={473},
  number={2202},
  pages={20160607},
  year={2017},
  publisher={The Royal Society},
  doi={10.1098/rspa.2016.0607}
}

@article{Wharton2020Colloquium,
  title = {Colloquium: {B}ell's theorem and locally mediated reformulations of quantum mechanics},
  author = {Wharton, K. B. and Argaman, N.},
  journal = {Rev. Mod. Phys.},
  volume = {92},
  issue = {2},
  pages = {021002},
  year = {2020},
  month = {May},
  publisher = {American Physical Society},
  doi = {10.1103/RevModPhys.92.021002}
}

@article{Frauchiger2018,
  title={Quantum theory cannot consistently describe the use of itself},
  author={Frauchiger, Daniela and Renner, Renato},
  journal={Nature Communications},
  volume={9},
  number={1},
  pages={3711},
  year={2018},
  publisher={Nature Publishing Group},
  doi={10.1038/s41467-018-05739-8}
}

@article{Bong2020NoGo,
  title={A strong no-go theorem on the {W}igner's friend paradox},
  author={Bong, Kok-Wei and Utreras-Alarc{\'o}n, An{\'i}bal and Ghafari, Farzad and Liang, Yeong-Cherng and Tischler, Nora and Cavalcanti, Eric G. and Pryde, Geoff J. and Wiseman, Howard M.},
  journal={Nature Physics},
  volume={16},
  number={12},
  pages={1199--1205},
  year={2020},
  publisher={Nature Publishing Group},
  doi={10.1038/s41567-020-0990-x}
}

@article{Badhan2026Standard,
  title={A Standard Quantum Mechanical Treatment to Rationalize the Delayed-Choice Quantum Eraser},
  author={Badhan, Vipul and {Pranvi} and Dhillon, Samreet and Arora, Bindiya},
  journal={arXiv preprint arXiv:2608.02185},
  year={2026},
  doi={10.48550/arXiv.2608.02185}
}

@article{Salart2008Speed,
  title={Testing the speed of `spooky action at a distance'},
  author={Salart, Daniel and Baas, Augustin and Branciard, Cyril and Gisin, Nicolas and Zbinden, Hugo},
  journal={Nature},
  volume={454},
  number={7206},
  pages={861--864},
  year={2008},
  publisher={Nature Publishing Group},
  doi={10.1038/nature07121}
}

@article{Bancal2012Finite,
  title={Quantum non-locality based on finite-speed causal influences leads to superluminal signalling},
  author={Bancal, Jean-Daniel and Pironio, Stefano and Ac{\'i}n, Antonio and Liang, Yeong-Cherng and Scarani, Valerio and Gisin, Nicolas},
  journal={Nature Physics},
  volume={8},
  number={12},
  pages={867--870},
  year={2012},
  publisher={Nature Publishing Group},
  doi={10.1038/nphys2460}
}

@article{Hubel2010Triplets,
  title={Direct generation of photon triplets using cascaded photon-pair sources},
  author={H{\"u}bel, Hannes and Hamel, Deny R. and Fedrizzi, Alessandro and Ramelow, Sven and Resch, Kevin J. and Jennewein, Thomas},
  journal={Nature},
  volume={466},
  number={7306},
  pages={601--603},
  year={2010},
  publisher={Nature Publishing Group},
  doi={10.1038/nature09175}
}

@article{Corona2011TOSPDC,
  title = {Third-order spontaneous parametric down-conversion in thin optical fibers as a photon-triplet source},
  author = {Corona, Mar{\'i}a and Garay-Palmett, Karina and U'Ren, Alfred B.},
  journal = {Phys. Rev. A},
  volume = {84},
  issue = {3},
  pages = {033823},
  year = {2011},
  month = {Sep},
  publisher = {American Physical Society},
  doi = {10.1103/PhysRevA.84.033823}
}

@article{Corcoles2021Dynamic,
  title = {Exploiting Dynamic Quantum Circuits in a Quantum Algorithm with Superconducting Qubits},
  author = {C\'orcoles, A. D. and Takita, Maika and Inoue, Ken and Lekuch, Scott and Minev, Zlatko K. and Chow, Jerry M. and Gambetta, Jay M.},
  journal = {Phys. Rev. Lett.},
  volume = {127},
  issue = {10},
  pages = {100501},
  year = {2021},
  month = {Aug},
  publisher = {American Physical Society},
  doi = {10.1103/PhysRevLett.127.100501}
}

@article{Jacques2007Wheeler,
  title={Experimental realization of {W}heeler's delayed-choice gedanken experiment},
  author={Jacques, Vincent and Wu, E and Grosshans, Fr{\'e}d{\'e}ric and Treussart, Fran{\c{c}}ois and Grangier, Philippe and Aspect, Alain and Roch, Jean-Fran{\c{c}}ois},
  journal={Science},
  volume={315},
  number={5814},
  pages={966--968},
  year={2007},
  publisher={American Association for the Advancement of Science},
  doi={10.1126/science.1136303}
}

@article{KimHam2023Coherent,
  title={Observations of the delayed-choice quantum eraser using coherent photons},
  author={Kim, Sangbae and Ham, Byoung S.},
  journal={Scientific Reports},
  volume={13},
  number={1},
  pages={9758},
  year={2023},
  publisher={Nature Publishing Group},
  doi={10.1038/s41598-023-36590-7}
}

@article{Deutsch1991CTC,
  title = {Quantum mechanics near closed timelike lines},
  author = {Deutsch, David},
  journal = {Phys. Rev. D},
  volume = {44},
  issue = {10},
  pages = {3197--3217},
  year = {1991},
  month = {Nov},
  publisher = {American Physical Society},
  doi = {10.1103/PhysRevD.44.3197}
}

@article{Lloyd2011CTC,
  title = {Closed Timelike Curves via Postselection: Theory and Experimental Test of Consistency},
  author = {Lloyd, Seth and Maccone, Lorenzo and Garcia-Patron, Raul and Giovannetti, Vittorio and Shikano, Yutaka and Pirandola, Stefano and Rozema, Lee A. and Darabi, Ardavan and Soudagar, Yasaman and Shalm, Lynden K. and Steinberg, Aephraim M.},
  journal = {Phys. Rev. Lett.},
  volume = {106},
  issue = {4},
  pages = {040403},
  year = {2011},
  month = {Jan},
  publisher = {American Physical Society},
  doi = {10.1103/PhysRevLett.106.040403}
}

@article{Hamel2014Triplets,
  title={Direct generation of three-photon polarization entanglement},
  author={Hamel, Deny R. and Shalm, Lynden K. and H{\"u}bel, Hannes and Miller, Aaron J. and Marsili, Francesco and Verma, Varun B. and Mirin, Richard P. and Nam, Sae Woo and Resch, Kevin J. and Jennewein, Thomas},
  journal={Nature Photonics},
  volume={8},
  number={10},
  pages={801--807},
  year={2014},
  publisher={Nature Publishing Group},
  doi={10.1038/nphoton.2014.218}
}

@article{Sen2020SuperdetI,
  title={Superdeterministic hidden-variables models {I}: non-equilibrium and signalling},
  author={Sen, Indrajit and Valentini, Antony},
  journal={Proceedings of the Royal Society A},
  volume={476},
  number={2243},
  pages={20200212},
  year={2020},
  publisher={The Royal Society},
  doi={10.1098/rspa.2020.0212}
}

@article{Sen2020SuperdetII,
  title={Superdeterministic hidden-variables models {II}: conspiracy},
  author={Sen, Indrajit and Valentini, Antony},
  journal={Proceedings of the Royal Society A},
  volume={476},
  number={2243},
  pages={20200214},
  year={2020},
  publisher={The Royal Society},
  doi={10.1098/rspa.2020.0214}
}

@article{Handsteiner2017Cosmic,
  title = {Cosmic {B}ell Test: Measurement Settings from {M}ilky {W}ay Stars},
  author = {Handsteiner, Johannes and Friedman, Andrew S. and Rauch, Dominik and Gallicchio, Jason and Liu, Bo and Hosp, Hannes and Kofler, Johannes and Bricher, David and Fink, Matthias and Leung, Calvin and Mark, Anthony and Nguyen, Hien T. and Sanders, Isabella and Steinlechner, Fabian and Ursin, Rupert and Wengerowsky, S\"oren and Guth, Alan H. and Kaiser, David I. and Scheidl, Thomas and Zeilinger, Anton},
  journal = {Phys. Rev. Lett.},
  volume = {118},
  issue = {6},
  pages = {060401},
  year = {2017},
  month = {Feb},
  publisher = {American Physical Society},
  doi = {10.1103/PhysRevLett.118.060401}
}

@article{Rauch2018Cosmic,
  title = {Cosmic {B}ell Test Using Random Measurement Settings from High-Redshift Quasars},
  author = {Rauch, Dominik and Handsteiner, Johannes and Hochrainer, Armin and Gallicchio, Jason and Friedman, Andrew S. and Leung, Calvin and Liu, Bo and Bulla, Lukas and Ecker, Sebastian and Steinlechner, Fabian and Ursin, Rupert and Hu, Beili and Leon, David and Benn, Chris and Ghedina, Adriano and Cecconi, Massimo and Guth, Alan H. and Kaiser, David I. and Scheidl, Thomas and Zeilinger, Anton},
  journal = {Phys. Rev. Lett.},
  volume = {121},
  issue = {8},
  pages = {080403},
  year = {2018},
  month = {Aug},
  publisher = {American Physical Society},
  doi = {10.1103/PhysRevLett.121.080403}
}

@book{tHooft2016CA,
  title={The Cellular Automaton Interpretation of Quantum Mechanics},
  author={{'t Hooft}, Gerard},
  series={Fundamental Theories of Physics},
  volume={185},
  year={2016},
  publisher={Springer},
  doi={10.1007/978-3-319-41285-6}
}

\appendix

\section{Fixed-point accounting for the feedback experiment}
\label{app:fixedpoint}

In this Appendix, we give the derivation of the results quoted in \Cref{sec:fixedpoint}. Let us denote the four types of hidden state using the pair $(u_\text{on}(\lambda), u_\text{off}(\lambda))$, with measures
\begin{equation}
\begin{split}
p_1 &:=\mu({\rm U}_4, {\rm U}_3), \quad p_2 := \mu({\rm U}_4, {\rm U}_4),\\
p_3 &:= \mu({\rm U}_3, {\rm U}_3), \quad p_4 := \mu({\rm U}_3, {\rm U}_4).
\end{split}
\end{equation}
The static constraints $\mu(u_\text{on} = {\rm U}_4) = p_1 + p_2 = 1/2$ and $\mu(u_\text{off} = {\rm U}_4) = p_2 + p_4 = 1/2$, together with normalisation, give
\begin{equation}
p_1 = p_4 = \frac{1}{2} - p_2, \quad p_3 = p_2, \quad p_2 \in \left[0, \tfrac{1}{2}\right],
\end{equation}
giving \Cref{eq:types}). Note the measure of paradox-admitting states, $p_4$, vanishes exactly when $p_2 = 1/2$, in which case $u_\text{on} = u_\text{off}$ almost everywhere.

For the feedback statistics, suppose the type-$({\rm U}_3,{\rm U}_4)$ states are excluded from the feedback ensemble (they admit no consistent history) and the ambiguous type-$({\rm U}_4,{\rm U}_3)$ states are resolved to the mirror-in history with probability $q$. Assume further that $d_\text{on}$ is independent of the type of $\lambda$ and of the resolution rule (the static constraints alone impose only independence of $d_\text{on}$ from $u_\text{on}$, which is weaker). The consistent measure is $p_1 + p_2 + p_3 = 1/2 + p_2$. For mirror-in histories, the upper outcome is U$_4$ and, by the static independence constraint, the lower outcome is D$_1$ with probability $1/2$ and otherwise D$_3$ or D$_4$ with probability $1/4$ each; for mirror-out histories, the upper outcome is U$_3$ and the perfect correlation constraint forces D$_3$. Therefore,
\begin{eqnarray}
P({\rm U}_3, {\rm D}_3) &=& \frac{p_3 + (1-q)\, p_1}{\tfrac{1}{2} + p_2} = \frac{p_2 + (1-q)\left(\tfrac{1}{2} - p_2\right)}{\tfrac{1}{2} + p_2}~, \nonumber\\
P({\rm U}_4, {\rm D}_1) &=& \frac{1}{2}\, \frac{p_2 + q \left(\tfrac{1}{2} - p_2\right)}{\tfrac{1}{2} + p_2}~,
\end{eqnarray}
with $P({\rm U}_4, {\rm D}_3) = P({\rm U}_4, {\rm D}_4) = P({\rm U}_4, {\rm D}_1)/2$. Setting $P({\rm U}_3, {\rm D}_3) = 1/2$ gives $q = 1/2$ whenever $p_2 < 1/2$, and with $q = 1/2$ all four joint probabilities take their quantum mechanical values $\left(\tfrac{1}{2}, \tfrac{1}{4}, \tfrac{1}{8}, \tfrac{1}{8}\right)$ for every $p_2$. For $p_2 = 1/2$ there are no ambiguous or paradoxical states, and no resolution rule is needed, so the quantum mechanical values are reproduced without any such rule---in this case, the independence of $d_\text{on}$ from the type of $\lambda$ reduces to the independence from $u_\text{on}$ which the static constraints already impose. Any $q \neq 1/2$ produces a measurable deviation, as does dropping the additional independence assumption: for example, at $p_2 = 1/4$, letting $d_\text{on} = {\rm D}_1$ on all of the $({\rm U}_4, {\rm U}_4)$ states, on none of the $({\rm U}_4, {\rm U}_3)$ states, and on half of the $u_\text{on} = {\rm U}_3$ states preserves both static constraints ($\mu(d_\text{on} = {\rm D}_1) = 1/2$ and $u_\text{on}$ independent of $d_\text{on}$), yet gives $P({\rm U}_4, {\rm D}_1) = 1/3$ with $q = 1/2$. A model with a live backwards channel ($p_2 < 1/2$) must therefore fine-tune both its resolution rule and the correlations among its response functions to remain compatible with quantum mechanics.

\end{document}